\documentclass[11pt]{article}

\usepackage{graphicx}
\usepackage{epstopdf}
\usepackage{amsmath}
\usepackage{natbib}
\usepackage{lipsum}    
\usepackage{amsmath}
\usepackage{amsmath,amssymb,amscd,enumerate,rotate,multirow}
\usepackage{bm,latexsym,mathrsfs}

\usepackage{amsthm}

\usepackage[hmargin=1.2in,vmargin=1.2in]{geometry}

\usepackage{xargs}                      

\usepackage[pdftex,dvipsnames]{xcolor}  
\usepackage{etoolbox}

\makeatletter
\patchcmd{\@makecaption}
  {\parbox}
  {\advance\@tempdima-\fontdimen2} 
  {}{}
\makeatother

\usepackage[colorinlistoftodos,prependcaption,textsize=tiny]{todonotes}
\newcommandx{\unsure}[2][1=]{\todo[linecolor=red,backgroundcolor=red!25,bordercolor=red,#1]{#2}}
\newcommandx{\change}[2][1=]{\todo[linecolor=blue,backgroundcolor=blue!25,bordercolor=blue,#1]{#2}}
\newcommandx{\info}[2][1=]{\todo[linecolor=OliveGreen,backgroundcolor=OliveGreen!25,bordercolor=OliveGreen,#1]{#2}}
\newcommandx{\improvement}[2][1=]{\todo[linecolor=Plum,backgroundcolor=Plum!25,bordercolor=Plum,#1]{#2}}
\newcommandx{\thiswillnotshow}[2][1=]{\todo[disable,#1]{#2}}
\newcommandx{\inline}[2][1=]{\todo[inline,size=\small,linecolor=Plum,backgroundcolor=orange!25,bordercolor=Plum,#1]{#2}}
\newtheorem{proposition}{Proposition}
\newtheorem{theorem}{Theorem}

\newtheorem{definition}{Definition}

\newtheorem{corollary}{Corollary}

\usepackage{graphicx,color,psfrag,pgfplots,tikz,xcolor}
 \usetikzlibrary{calc,decorations.markings,positioning,shapes,positioning,arrows,patterns,fadings,shapes.arrows,shadows}
 \tikzstyle{every picture}+=[remember picture]
 \tikzset{arrowfill/.style={top color=gray!50, bottom color=gray!50, general shadow={fill=black}}}
\tikzset{arrowstyle/.style={draw=gray!50,arrowfill, single arrow,minimum height=#1, single arrow,
single arrow head extend=.15cm,line width=0.1mm,}}

\tikzstyle{abstractbox} = [draw=gray, fill=gray, rectangle,inner sep=8pt, style=rounded corners]
\tikzstyle{abstracttitle}=[fill=gray]

\usetikzlibrary{shadows,arrows,positioning}
\pgfdeclarelayer{background}
\pgfdeclarelayer{foreground}
\pgfsetlayers{background,main,foreground}

\tikzstyle{materia}=[draw, fill=white, text width=5cm,
  minimum height=2.5em]  
\tikzstyle{etape} = [materia, text width=17cm, minimum width=5cm,
  minimum height=4.em, rounded corners]  
\tikzstyle{decision} = [materia, text width=10.5cm, minimum width=5cm,
  minimum height=4.5em, rounded corners]  
  \tikzstyle{startstop} = [rectangle, rounded corners, minimum width=4.4cm, rounded corners,minimum height=0.7cm,
draw=black,minimum height=3.em, fill=blue!20!white]
\tikzstyle{process} = [rectangle, minimum height=4.em, rounded corners, minimum width=6.3cm, rounded corners,draw=black, fill=blue!20!white]
\tikzstyle{texto} = [above, text width=6em, text centered]
\tikzstyle{linepart} = [draw, thick, color=black!50, -latex', dashed]
\tikzstyle{line} = [draw, thick, color=black!50, -latex']
\tikzstyle{ur}=[draw, text centered, minimum height=0.01em, draw=white]

\newcommand{\etape}[2]{node (p#1) [etape]
  {#2}}
\newcommand{\start}[2]{node (p#1) [startstop]
  {#2}}

\newcommand{\decision}[2]{node (p#1) [decision]
  {#2}}

\newcommand{\be}{\begin{equation}}
\newcommand{\ee}{\end{equation}}

\newcommand{\iid}{\stackrel{\mathrm{iid}}{\sim}}
\newcommand{\ind}{\stackrel{\mathrm{ind}}{\sim}}
\newcommand{\e}{\mathrm{e}}
\newcommand{\calM}{\mathcal{M}}
\newcommand{\calB}{\mathcal{B}}

\newcommand{\calL}{\mathcal{L}}
\newcommand{\calN}{\mathcal{N}}
\newcommand{\calP}{\mathcal{P}}

\newcommand{\calX}{\mathcal{X}}

\newcommand{\Prob}{\mathbb{P}}
\newcommand{\E}{\mathbb{E}}
\newcommand{\ff}{\ _2F_1 }
\newcommand{\LL}{\mathcal L}

\newcommand{\Rea}{{\mathbb{R}}}

\newcommand{\uno}{\mathbb{I}}

\DeclareMathOperator{\Var}{Var}

\usepackage[normalem]{ulem}

\begin{document}

\title{\vspace{-2cm} \bf Is infinity that far?\\A  Bayesian nonparametric perspective of finite mixture models}
  \author{Raffaele Argiento\hspace{.2cm}\\
    \small{ESOMAS Department, University of Torino and Collegio Carlo Alberto, Torino, Italy} \\
    and \\
    Maria De Iorio\\
    \small{Yale-NUS College, Singapore and Dept. of Statistical Science, University College London}
  }
    \date{}
    \maketitle

\begin{abstract}
Mixture models are  one of the most widely used statistical tools when dealing with data from heterogeneous populations. This paper considers the long-standing debate over finite mixture and infinite mixtures and brings the two modelling strategies together,  by showing that a finite mixture is simply a realization of a point process. Following a Bayesian nonparametric perspective, we introduce a new class of prior: the Normalized Independent Point Processes. We investigate the probabilistic properties of this new class. Moreover, we design a conditional algorithm for finite mixture models with a random number of components overcoming the challenges associated with the Reversible Jump scheme and the recently proposed marginal algorithms. 
We illustrate our model on real data and discuss an important application  in population genetics.
\end{abstract}
\section{Introduction}
\label{sec:intro}
Mixture models are a very powerful and natural statistical tool  to model data  
from heterogeneous populations.   In a mixture model, observations  are  assumed to have arisen from one of $M$ (finite or infinite) groups, each group being suitably modelled by a density typically from a  parametric family. The density of each group is referred to as a component of the mixture, and is weighted by the relative frequency (weight) of the group in the population. This model offers a conceptually simple  way of relaxing distributional assumptions and 
 a convenient and flexible  way to approximate distributions that cannot be modelled satisfactorily by a standard parametric family. Moreover, it provides a framework by which observations may be clustered together into groups for discrimination or classification. For a comprehensive review of mixture models and their  applications see \cite{mclachlan2000finite, fruhwirth2006finite} and \cite{fruhwirth2019handbook}.  More in details, let $Y\in\mathcal Y\subset \Rea^{r}$ be the population variable, each observation is assumed to have arisen from one of $0<M\le\infty$ groups: 
\begin{equation}
  \label{eq:mixture1}
  f_{Y}(y \mid P)=\int_{\Theta}^{}f(y \mid \theta)P(d\theta)=\sum_{j=1}^{M}w_j
  f(y\mid \tau_j)
\end{equation}
where $\{f(\cdot\mid\theta),\theta\in \Theta\subset \Rea^{d}\}$ is a parametric family of densities on $\mathcal Y$, while $P(\cdot)$ is an almost sure discrete measure on $\Theta$, and it is referred to as \emph{mixing measure}. Here $\{\tau_j, j=1,\ldots, M\}$ is a collection of points in  $\Theta$, that defines the support of $P$. For each $j=1,\dots,M$, the density  $f(y\mid \tau_j)$ is the kernel of the mixture, and is weighted by $w_j$, the relative frequency of the group in the population. Model \eqref{eq:mixture1} defines a framework by which observations may be  clustered together into groups, so that conditionally, data are independent and identically  distributed within the groups and independent between groups. To avoid confusion in terminology, in what follows  $M$ will denote the number of components in a mixture, i.e. of possible clusters/sub-populations, while by number of clusters, $k$, we mean the number of allocated components, i.e. components to which at least one observation has been assigned. The latter quantity can only be estimated \textit{a posteriori}.  

We believe that in the context  of mixture modelling the words cluster and component are often  misused in terminology, i.e. the distinction between number of components and number of clusters has generally been overlooked in the parametric world, leading to the criticism that if we fix \textit{a priori} the number $M$, we cannot estimate the number of clusters. What needs to be highlighted  \citep[see][]{rousseau2011asymptotic} is that when in a finite mixture model we fix $M$, we are specifying the number of components (i.e. possible clusters) that corresponds to the data generating process, but still we need to estimate the actual number of clusters in the sample (allocated components).   Already \cite{nobile2004posterior} had pointed out this difference, noticing that the posterior distribution of the number of components $M$ might assigns considerable probabilities to values greater than the number of allocated components. 
Similar observations have been by \cite{richgreen97}, who specify a prior on the number of components $M$, highlighting the fact that some of the components might be empty as not all the components might be represented in a finite sample and the data are non-informative on unallocated components. This leads to an identifiability problem for $M$ and, as a consequence, fully non-informative priors cannot be elicited in a mixture context. Nevertheless, \cite{richgreen97} still focus their inference problem on $M$ and do not investigate the relationship between $M$ and $k$.  More recently, \cite{malsiner2016model} introduce sparse finite mixture models as an alternative to infinite (nonparametric) mixtures, and impose sparsity to estimate the number of non-empty components in a deliberately over-fitting mixture model where $M$ is fixed relatively large. On the other hand, in Bayesian nonparametrics $M$ is set equal to infinity  (i.e., $M=\infty$) and the focus of inference is only $k$. In this work we stress the importance of the distinction  between $M$ and $k$ as it will allow us to  collocate nonparametric and parametric mixtures in exactly the same  framework. 

 In Bayesian Nonparametrics  (i.e., $M=\infty$), the Dirichlet process mixture model \citep[DPM,][]{lo84,neal2000markov} -- i.e. the Model in \eqref{eq:mixture1} where the mixing measure is indeed  the Dirichlet  process --  plays a pivotal role. DPM popularity is mainly due to its high flexibility and mathematical and computational  tractability  both in density estimation as well as in clustering problems.
However, in some statistical applications, the use of the Dirichlet process as a clustering mechanism may be restrictive \citep[see, for instance,][]{lau2007bayesian,miller2013simple}:  the clustering results often depend on the choice of a particular kernel, partitions will typically be dominated by few  large clusters (the \textit{rich-get-richer} property), the number of clusters increases as the  number of observations $n$ increases (as $\log n$), often leading to the creation  of too many non-interpretable singleton clusters.  To overcome these drawbacks many alternative mixing measures have been proposed \citep[e.g.][]{ishwaran2003some,dey2012practical}. In particular, \cite{controlling}  replace the Dirichlet process with a large and flexible class of random probability measures obtained by \emph{normalization} of random (infinite dimensional) measures with independent increments. Once again, all these approaches assume $M=\infty$ and focus on estimating $k$.

On the other hand, in a Bayesian parametric context (i.e., $M<\infty$ almost surely) the most popular approaches are (i) fix $M$ and then focus mainly on density estimation (ii) treat   $M$ as a random parameter and make it the focus of inference. Then, conditionally on $M=m$, the mixture weights $(w_1,\dots,w_m)$ are chosen according to a $m-1$ dimensional $Dirichlet_m$ distribution. We refer to the latter model as \emph{finite Dirichlet mixture model} (FDMM). Refer,  among the others, to \cite{nobile94,richgreen97,steph00} and \cite{miller2017mixture} for more details. The literature is rich of proposals on how to estimate the number of components $M$, but there is no consensus on the best method. Likelihood based inference typically relies on model choice criteria, such as BIC  or the approximate weight of evidence \citep[see][for a review]{biernacki2000assessing}. Although in the Bayesian paradigm there are approaches based on model choice criteria, such as DIC, it is usually preferable to perform full posterior inference on $M$ as well, eliciting an appropriate prior. A fully Bayesian approach     in FDMM is often based on the reversible jump Markov chain Monte Carlo \citep{richgreen97,dellaportas2006multivariate} or, alternatively, on the marginal likelihood $p(\boldsymbol{y}\mid M)$. Both methods present significant computational challenges.   

Although mainly for computational purposes, the connection between finite and infinite mixture models has been present in the literature for at least two decades since the  work of \cite{muliere1998approximating}. Since then extensive research effort has been devoted to find approximate representation of the Dirichlet process \citep[e.g.][]{ishwaran2002exact}. Moreover,   algorithms for posterior inference of infinite mixture models often truncate the infinite measure  and approximate it with a finite mixture with $L$ components, where $L$ is sufficiently large or random \citep[see for instance][]{ishwaran2001gibbs,argiento2016} and, in practice, the inferential problem translates  into estimating the number of allocated components (clusters) and the cluster-specific parameters.  The focus of this work is to provide a  probabilistic treatment of mixture modelling, that reconciles the two approaches: $M=\infty $ and $M< \infty $.  
This connection has received some attention from a theoretical point of view \citep[see e.g.][]{miller2017mixture,fruhwirth2017here}, but has never been investigated thoroughly  and fully resolved.
  
\subsection{Contribution of this work}   
  
In this work, instead of approximating an infinite mixture with a parametric one, we show that a finite  mixture model is simply a realization of a  stochastic process whose dimension is random and has an infinite dimensional support.  To this end, we introduce  a new class of random measures obtained by normalization of a point process and use it as mixing measures in Model \eqref{eq:mixture1}. We refer to this new class  as \emph{Normalized Independent Finite Point Processes} and we derive  the family of prior distributions induced on the data partition by providing  a general formula for  exchangeable partition probability functions \citep{pitman96}. Finally, we characterize the posterior distribution of the  Normalized Independent Finite Point Process.  
Our construction is exactly in the spirit of Bayesian nonparametrics, as it is based on the  normalization of a  point process, leading to an almost surely discrete measure. Indeed,  there is a fundamental and simple idea behind the construction of almost surely discrete random measures: they can be obtained by normalization of stochastic processes. Already \cite{ferguson1973bayesian} in his seminal work derived the Dirichlet process as normalization of a Gamma process. More recently, \cite{regazzini2003distributional}  propose a new class of nonparametric priors, called \textit{Normalized Random Measures with Independent Increments}, obtained through the normalization of a L\'evy process.   This latter work has opened  the door to one of  the most active    lines of current research in Bayesian statistics as well as in machine learning. On one hand it has led to the development of nonparametric priors beyond the Dirichlet process \citep[e.g.][]{lijoi2010models} and on the other the same techniques are widely used for clustering  in the machine learning community under the name of \textit{Normalised Completely Random  Measures } \citep[e.g.][]{jordan2010hierarchical}.

The class we propose is rich and includes as a particular case the popular finite Dirichlet  mixture model. Several inference methods have been  proposed for the finite  mixture models, of which the most commonly-used are the  Reversible Jump Markov chain Monte Carlo \citep{green1995reversible,richgreen97} and the recently proposed marginal algorithm \citep{miller2017mixture}. The Reversible jump algorithm is a very general technique and has been successfully applied in many contexts, but it can be difficult to implement since it requires  designing problem-specific moves, which is often a nontrivial task particularly in high-dimensions. The algorithm proposed by \cite{miller2017mixture}, although more efficient and in some ways more automatic, restrains the class of prior distributions for the weights and does not allow  inference on the hyper-parameters of the process, which could constitute a serious limitation in complex set-ups. In \cite{miller2017mixture}, by integrating out the mixing measure,  inference is limited to the number of allocated components and  the  mean of linear functionals of the posterior distribution of the mixture model. See \cite{gelfand2002computational} for a discussion of these issues.

Among the  main achievements of this work, there is the construction of a Gibbs sampler scheme to simulate from the posterior distribution of the Normalized Finite Independent Point Process, in particular a  conditional MCMC algorithm  based on the posterior characterization of such process. This algorithm, in the particular case of a Dirichlet prior on the mixture weights,  leads to conjugate updating with a substantial gain in computational efficiency over current algorithms.  The key result (associated to the nonparametric construction of the process)  is to be able to propose transdimensional moves which are automatic and naturally implied by the prior process.  We illustrate the proposed prior process through the benchmark example offered by the  the Galaxy data \citep{roeder1990density} and an important application in population genetics.

The manuscript is structured as follows. In Section~\ref{sec:fmm} we introduce the finite mixture model framework, highlighting the connection between parametric and nonparametric constructions.  Section~\ref{sec:ffp} reviews necessary theory from Finite Point Processes. In Section~\ref{sec:nifpp} we introduce the prior process, the Normalised Independent Finite Point Process, and discuss its clustering properties, while in Section~\ref{sec:postnorm} we characterise its posterior distribution. In Section~\ref{sec:posterior_inference} we show how the new construction leads to efficient conditional and marginal algorithms. In Section~\ref{sec:nfpd} we briefly show how the new prior can be used as a component in more complex hierarchies. In Section~\ref{sec:special_q} we discuss possible choices for the prior on the number of mixture components, while Section~\ref{sec:impex}
presents special cases of the process. We conclude the paper with two  examples: (i) the Galaxy data example, which provides an opportunity to benchmark our method  in Section~\ref{sec:Galaxy}; (ii) a real data genetic application aimed to identify population structure from microsatellites loci genotyped in a sample of thrushes in Section~\ref{sec:popstruct}.  We conclude the paper in Section~\ref{Conclusions}.

\section{Finte Mixture Models} 
\label{sec:fmm}

In this section we introduce the finite mixture model (FMM) and show how it can be written in three equivalent ways. The first two representations are widely used in  the parametric literature, while the last one uses notions typical of  random mixing measure (nonparametric) set-ups. Our starting point is the general finite mixture model. Let $Y_1,\dots,Y_n$ be a set of observations taking values  in an Euclidean space $\mathcal Y$. We  consider the following mixture model:
\begin{equation}
\begin{split}
& Y_i| \tau_1, \ldots, \tau_M, \mathbf{w}, M \iid \sum_{m=1}^M w_m f(y\mid \bold\tau_m)\\
& \tau_i\mid M  \iid P_0(d\tau), \qquad i=1,\ldots, M  \\
& \mathbf{w} \mid M \sim P_W(\mathbf{w})\\
& M\sim q_M
\end{split}
 \label{eq:FMM}
\end{equation}
where $f(\cdot;\tau_m)$ is a parametric  density on $\mathcal{Y}$, which depends on a vector of parameters $\tau_m$. The vector of parameters assume values in  $\Theta \subset \Rea^d$ and is assigned a non-atomic prior density  $p_0$ corresponding to the probability measure $P_0$ on $\Theta$. 
The number of components is an important parameter of the mixture model and in a fully Bayesian approach it is given a prior $q_M$. Conditionally on $M$, the vector of weights $\mathbf{w}=(w_1,\ldots, w_M)$, which represents the probability of belonging to each mixture component, is given a prior probability $P_W$ on the simplex of dimension $M-1$. The model in Eq. \eqref{eq:FMM} can be rewritten in terms of latent  variables, since this representation allows for simpler computations. To this end, we need to introduce a latent allocation vector $\mathbf{c} = (c_1,\ldots, c_n) $, whose element $c_i$ denotes to which component observation $Y_i$ is assigned, $c_i \in \{1, \ldots, M\} $. Then the model in \eqref{eq:FMM} is equivalent to:
\begin{equation}
\begin{split}
  & Y_i| \theta_i \ind  f(y\mid \theta_i), \qquad i=1,\dots,n\\
  & \theta_i|c_i\ind \delta_{\tau_{c_i}}(d\theta_i)\\
& \tau_m\mid M  \iid P_0(d\tau), \qquad m=1,\ldots, M  \\
& \mathbf{w} \mid M \sim 
P_{W}(\bm w\mid M)\\
&c_i\mid M, \mathbf{w} \sim  \text{Multinomial}_M (1, w_1,\ldots,w_M)  \\
& M\sim q_M
\end{split}
 \label{eq:latvar}
\end{equation}
where $\delta_\tau$ is  the Dirac measure assigning unit mass at location $\tau$.
Usually $P_W$ is assumed to be a Dirichlet$_M(\gamma,\ldots, \gamma)$ distribution, while typical choices for $q_M$ include a discrete uniform on some finite space, a Negative Binomial or a Poisson distribution. 
Note that prior information about the relative sizes of the mixing weights $w_1,\ldots,w_M$ can be introduced through $\gamma$ -- roughly speaking, small $\gamma$  favours lower entropy $w$'s, while large $\gamma$ favours higher entropy $w$'s. In general, 
 the hyperparameter $\gamma$ is  either set equal to a constant (e.g. equal 1 or $1/M$)  or is assigned a Gamma hyperprior.  
\cite{fruhwirth2017here} propose a sparsity prior on $\mathbf{w}$, which allows the number of non-empty components to be much smaller than $M$, where  $M$ is a non-random constant.
Alternatively, \cite{miller2017mixture} do not make strong assumptions on the prior on $M$ but, by showing the connection between FMM and exchangeable partition probability functions (eppf), manage to apply the well-developed inferential methods  for DPMs  to FMMs with significant gains in computational efficiency. The strategy proposed by  \cite{miller2017mixture} is limited to the Dirichlet prior on $\mathbf{w}$ and employs a marginal-type of algorithm to perform posterior inference. This approach, often used in DPMs, marginalises over the weights of the mixture and it is most appropriate when the main object of scientific interest is $k$.
In this work we propose a richer construction, where the prior on $\mathbf{w}$ is obtained normalising a finite point process. Advantages of the proposed approach include: (i) extension of the family of prior distributions for the weights; (ii) full Bayesian inference on all the unknowns (in particular $M$ and $\mathbf{w}$); (iii) possibility of inducing sparsity through appropriate choice of hyper-parameters; (iv) ease of interpretation; (v) possibility of extending the construction to covariate dependent weights and (vi) extension to more general processes.

In a nutshell, we build a general class of finite mixture models by proposing a new prior process for 
$P_W, q_M, P_0$ which admits the conventional mixture model described in Eq.~\eqref{eq:latvar} as special case. To introduce this new class of prior distributions,  which we refer to as {\em Normalized Independent Finite Point Processes} (Norm-IFPP), we first need to review some background theory and introduce some notation. Then, in Section \ref{sec:impex} we provide examples which do not require the prior for the weights to be a Dirichlet distribution.

The theoretical developments are based on the key observation that a realization  $M,\bm w,\bm \tau$  from  the prior on the mixture model parameters defined in Eq.~\eqref{eq:FMM} in terms of  hierarchical parametric distribution $q_M, P_W, P_0$ defines an almost surely (a.s.) finite-dimensional random probability measure on the parameter space $\Theta$:
\begin{equation}
P(\theta) = \sum_{m=1}^M w_m \delta_{\tau_m}(d\theta)
\label{eq:P}
\end{equation}	 
This implies that the joint probability distribution on $M$, $\mathbf{w}$ and $\boldsymbol{\tau}$ induces a distribution on $P$ defined in Eq.~\eqref{eq:P}, whose support is  the space of the a.s. finite-dimensional random probability measures on $\Theta$. Moreover, it is straightforward  to prove \citep[see][]{Arg_etal18_sub} that by letting $\theta_i=\tau_{c_i}$, as in Eq.~\eqref{eq:latvar}, the variables $\theta_1,\dots,\theta_n$ can be considered as a sample from $P$, i.e. $\theta_1,\dots,\theta_n|P\iid P$. From this observation,  the link between infinite (nonparametric) and finite mixture models becomes evident as the model in Eq.~\eqref{eq:FMM} can be easily rewritten as
\begin{equation}
  \begin{split}
    &Y_1,\dots,Y_n |\theta_1,\dots,\theta_n \stackrel{ind}{\sim}
    f(y;\theta_i)\\
    &\theta_1,\dots,\theta_n |P \stackrel{iid}{\sim}P \\
  &   P\sim   \mathcal{P}  
  \end{split}
  \label{eq:modello}
\end{equation}
where $P$ is defined in Eq.~\eqref{eq:P} and $\mathcal{P}$ is the law of $P$ defined via $q_M,P_W,P_0$.  The main theoretical contribution of this work is to give a constructive definition of $\mathcal{P}$, introducing a class of FMM for which the weights $w_m$ represent the normalised jumps of a \emph{finite point process}  and  the parameters $\tau_m$ are defined in terms of   realisations of the same point process. As in any mixture, $\theta_i$s in Model~\eqref{eq:modello} are equal to one of the $\tau_m$ in Eq.~\eqref{eq:P}, depending on which component the $i$th observation is assigned to.  
The link between finite mixture models and point processes is not unknown, as pointed out in the introduction. 
In particular, \cite{steph00} highlights this connection, but defines the point process on the complex space of normalized weights (i.e. the union of infinite simplexes). In this work through normalization  not only we are able to work on a simpler space, but also to build a new general class of distribution $P_W$, i.e. a new class of prior for the weights of the mixing measure $\bm w$.

\section{Finite Point Processes}
\label{sec:ffp}
In this section we review some concepts from point process theory which are necessary to construct the Norm-IFPP process. We refer to the  books of \cite{daleyverej} and \cite{moller2003statistical} for a complete treatment of finite point processes.

Let $\mathcal X$ be a complete separable metric space, a \emph{finite point process}  $X$ is  a random countable subset of $\mathcal X$. In this paper we restrict our  attention  to processes whose realizations are a finite subset of $\mathcal X$. For any realization of the process, $x\subseteq \mathcal X$, let $\# (x)$ denote the cardinality of $x$. The realizations of $X$ are constrained on $N_{f}=\{x\subseteq \calX :\# (x)<\infty \}.$
Elements of $N_f$ are called \emph{finite point configurations}.  The law of a finite point process is identified by the following quantities: 
 \begin{enumerate}
     \it{
    \item a discrete probability density $q_M, M=0,1,\dots$ 
       determining the law  of the total number $M$ (i.e. $\# (x)$)  
      of points of the process,
    \item for each integer $M \ge 1$, a probability distribution $\Pi_M(\cdot)$ 
      on the Borel sets of $\mathcal{X}^{M}$, that determines the joint distribution of
      the positions of the points of the process, given that their total number is
    $M$.}
  \end{enumerate}
In particular, $q_M$ and $\Pi_M$ provide
a constructive definition of the process which is very useful in simulations. First
generate a random integer $M$ from $q_M$ and then, given $M \neq 0$,
generate a random set $X=\{\xi_1 ,\dots, \xi_M\}$ which is a sample from  $\Pi_M(\cdot)$.
If $M=0$, the random generation stops and $X$ coincides with the empty set.

Note that a point process  $X=\{\xi_1,\dots,\xi_M\}$ is a set of
unordered points. 
To this end, the distributions $\Pi_M(\cdot)$ needs to give equal weight to
all $M!$ permutations of the elements in the vector $(\xi_1 ,\dots, \xi_M)$,
i.e. $\Pi_M (\cdot)$ must be symmetric.
A convenient way to specify the law of $X$ is based on the \emph{Janossy measure} \citep{daleyverej}:
 \begin{equation*}
  J(A_1 \times \dots \times A_M ) 
  = M! q_M \Pi_M(A_1\times\dots\times A_M)
\end{equation*}
where the $A_m$s are Borel-sets of $\mathcal{X}$, with $\xi_m \in A_m$.
The \emph{Janossy measure} is  unnormalised  and plays a  fundamental role in the
study of finite point processes and  spatial point patterns.
It has a simple interpretation which makes it easy to work with.   
Let  $\mathcal X = \Rea^d$ and let $j(\xi_1,\dots, \xi_M )$ denote the density of
$J(\cdot)$ with respect to the Lebesgue measure  with
$\xi_m\neq \xi_{\tilde{m}}$ for $m\neq \tilde{m}$. Then 
  $j(\xi_1,\dots,\xi_M)d\xi_1\dots d\xi_M$ is the probability that there are exactly $M$ points
    in the process, one in each of the distinct infinitesimal regions $
  (\xi_m,\xi_m+d\xi_m) $.
Here, we will use the Janossy measure to characterize the prior and the posterior distribution of the  new class of finite discrete
random probability measures, i.e. the family of \emph{Normalised Independent Finite Point Processes} (Norm-IFPP). We now introduce a simplified version of this process which assumes that the points $\xi_m$ are conditionally independent and identically distributed.  
\begin{definition}
  Let $\nu(\cdot)$ and  $\{q_M, M=0,1,\dots\}$ be a density on $\calX$ and a 
  probability mass function respectively. $X$ is an 
  {\bf independent finite point process},  $X\sim IFPP(\nu,q_M)$,
  if its Janossy density can be written as 
  \begin{equation}
  \label{eq:ijanossy}
    j(\xi_1,\dots,\xi_M) = M!q_M\prod_{m=1}^{M}\nu(\xi_m)
  \end{equation}
\end{definition}
In what follows, our construction is based on Eq.~\eqref{eq:ijanossy}. 

\section{Normalized Independent Finite Point Processes}
\label{sec:nifpp}
Let $\Theta\subset \Rea^d$, for some positive integer $d$ and let $\calX$ be  $\Rea^{+}\times\Theta$. We denote with  $\xi=(s,\tau)$ a point of $\calX$. Let   $\nu(s,\tau)$ be a density on $\calX$ such that $\nu(s,\tau)=h(s)p_{0}(\tau)$, where $h(\cdot)$ is a density on $\Rea^{+}$ and $p_0(\cdot)$ is a density on $\Theta$.  Finally, we consider only $q_M$ such that $q_0=0$, i.e. the prior probability of $M=0$ is zero. 
We consider the   independent finite point process $\tilde P=\{(S_1,\tau_1),\dots,(S_M,\tau_M)\}$ with parameters $\nu$ and $\{q_M\}$, i.e. $\tilde P \sim IFPP(\nu,q_M)$. In what follows, it is easier to introduce a slight change of notation and define $IFPP(h,q_m,p_0) = IFPP(\nu,q_M)$ to highlight the dependence of the process also on $p_0(\cdot)$. Let $\mathcal M:=\{1,\dots,M\}$ be the set of indexes corresponding to the points of the process. Since we are assuming that $q_0=0$ the random variable $T:=\sum_{m\in \mathcal M}^{}S_m$ is almost surely larger than $0$  so that we can give the following definition:
\begin{definition}
  Let $\tilde P=\{(S_1,\tau_1),\dots,(S_M,\tau_M)\}\sim IFPP(h,q_m,p_0)$, with $q_0=0$.
  A normalized independent finite point process (Norm-IFPP) with parameters $h(\cdot),\ p_0$ and $\{q_M\}$ is a discrete probability measure on $\Theta$ defined by
  \begin{equation}
    \begin{split}  
    P(A)&=  
    \sum_{m \in \mathcal M}^{}w_m\delta_{\tau_m}(A) 
    \stackrel{d}{=}\sum_{m \in \mathcal M}^{
    }\frac{S_m}{T}\delta_{\tau_m}(A)
  \end{split}    
    \label{eq:nor_ifpp}
  \end{equation}
  where $T=\sum_{m\in \mathcal M}^{}S_m$ and $A$ denotes a measurable set of $\Theta$. We  refer to the process in Eq.~\eqref{eq:nor_ifpp} as $P\sim
  {\rm Norm-IFPP} (h,q_m,p_0)$.
\end{definition}
The finite dimensional process defined in Eq.~\eqref{eq:nor_ifpp} belongs to the wide class of species sampling models \citep[see][]{pitman96} and this will allow us to use all the efficient machinery developed for such models.  Let $\left(\theta_1,\dots,\theta_n\right)$ be a sample from a Norm-IFPP. It is well known that sampling  from a discrete  probability measure induces ties among the $\theta_i$s and, therefore, a random partition of the observations. Let ${ \rho}_n:=\{C_1,\dots,C_k\}$ indicate a partition of the set $\{1,\dots,n\}$ in $k$ subsets, where $C_j = \{i : \theta_i = \theta^\star_j\} $ for $j=1,\dots,k\leq n$, and let $\{\theta_1^\star,\ldots, \theta_k^\star\}$ denote the set of distinct $\theta_i$s associated to each $C_i$. The marginal law of $\left(\theta_1,\dots,\theta_n\right)$ has a unique characterization: 
$$
\LL\left(d\theta_1,\dots,d\theta_n\right)=
\LL({ \rho}_n,d\theta^\star_1,\dots,d\theta^\star_k)=\pi(n_1,\dots,n_k) \prod_{j=1}^{k}P_0(d\theta^\star_j)$$ 
where $n_j=\#(C_j)$, $\sum_{j=1}^k n_j = n$ and $\pi(\cdot)$ is the exchangeable partition probability function (eppf) associated to the random probability $P$ \citep{pitman96}. For each $n$, the eppf $\pi$ is a probability law on the set of the partitions of $\{1,\dots,n\}$, which determines the (random) number of clusters $k$  and the numerosity of each cluster $C_i$. The partition is exchangeable because its law depends only on the number and size of the clusters, and not on the allocation of the individuals to each clusters.  
The eppf is a key tool in Bayesian analysis as mixture models can be rewritten in terms of random partitions and such equivalence  is often exploited to improve computational efficiency, in particular of marginal algorithms \citep{lijoi2010models}. The following  proposition provides an expression for the eppf of a Norm-IFPP measure. 
\begin{theorem}\label{thm:eppf_peps}
  Let  $(n_1,\dots,n_k)$ be a vector of positive integers such that $\sum_{j=1}^{k}n_j=n$. Then, the eppf 
  associated with  a Norm-IFPP$(h,p_0,q_M)$ is 
  \begin{equation}
    \pi(n_1,\dots,n_k)  =  
    \int_{0}^{+\infty}
    \dfrac{u^{n-1}}{\Gamma(n)}
    \left\{ 
      \sum_{m=0}^{\infty}\frac{(m+k)!}{m!}\psi(u)^{m}q_{m+k}
    \right\}
    \prod_{j=1}^k 
    \kappa(n_j,u)du  
    \label{eq:eppf_expr}
  \end{equation}
  where $\psi(u)$ is the Laplace transform of the density $h(s)$, i.e.
  \begin{equation}
    \label{eq:psi}
    \psi(u):=\int_{0}^{\infty}e^{-u s}h(s)ds
\end{equation}
and 
\begin{equation*}  
    \kappa(n_j,u):=
    \int_{0}^{\infty}s^{n_j}e^{-u s}h(s)ds=
    (-1)^{n_j}\frac{d}{du^{n_j}}\psi(u) 
  \end{equation*}
\end{theorem}

\noindent \textit{Proof:} See Appendix $\blacksquare$

For what follows it is important to highlight the difference between $M$ and $k$. The number of components of the finite mixture $M$ is given by a realisation of the process in Eq.~\eqref{eq:nor_ifpp}. On the other hand, $k$ denotes the number of  non-empty (allocated) components, with $k\leq M$. This difference has been noted before in the literature (see, for example, \cite{nobile2004posterior,miller2017mixture,  fruhwirth2017here}). Suppose that a realization from $\mathcal{P}$ is a discrete measure with $M=4$ atoms in Eq.~\eqref{eq:P} and $\boldsymbol{\tau} = (0.2, 2.4,1.5,4.1)$. Furthermore we have a realization from $P$, $\boldsymbol{\theta} = (0.2, 0.2, 4.1, 2.4,0.2,2.4)$, with $n=6$. Then the allocated components are $k=3$ and the total number of mixture components is $M=4$. 
Note that the representation in Eq.~\eqref{eq:nor_ifpp} implies that the jumps of the point process, indexed by the elements of $\mathcal{M}=\{1,\dots,M\}$, correspond to the components of the finite mixture, and their relative size defines the weights. More formally, we denote by  $\mathcal M^{(a)}$ the set of indexes of \emph{allocated} jumps of the Process~\eqref{eq:nor_ifpp}, i.e.  the indexes $m \in \mathcal M$ corresponding to some jumps $S_m$ such that there exists a location for which $\tau_{m}=\theta^\star_i$, $i=1, \dots, k$. The remaining values of $\mathcal M$ correspond to the \emph{non-allocated} jumps and we denote this set with $\mathcal M^{(na)}$. We use the superscripts $(a)$ and  $(na)$ for random variables related to \emph{allocated} and \emph{non-allocated} jumps respectively. 

One of the main focus of inference when using finite mixture models is to determine the clustering allocation of the observations. The eppf gives the prior distribution on the space of possible partitions.  Moreover, marginalising over the cluster sizes, it is also possible  to derive the implied prior distribution on the number of clusters, $k$, which corresponds to the number of allocated components. 

\begin{corollary}
\label{cor:priorK}
  Under the assumptions of Theorem~\ref{thm:eppf_peps}, the marginal prior probability of sampling a partition with $k$ clusters is given by 
\begin{equation}
\label{eq:priorK}
p^\star_k=\Pr\{M^{(a)}=k\}
=
   \int_{0}^{+\infty}
    \dfrac{u^{n-1}}{\Gamma(n)}
    \left\{ 
      \sum_{m=0}^{\infty}\frac{(m+k)!}{m!}\psi(u)^{m}q_{m+k}
    \right\}
    B_{n,k}(\kappa(\cdot,u))
 \end{equation} 
 where $k=1,\ldots,n$, and $B_{n,k}(\kappa(\cdot,u))$ is the partial Bell polynomial \citep{pitman2006combinatorial} over the sequence of coefficients $\{\kappa(n,u),\ n=1,2,\dots\}$.

\end{corollary}

\noindent \textit{Proof:} See Appendix~\ref{sec:app_proof_priorK} $\blacksquare$

Moreover, from  de Finetti's theorem it follows that, $k$ converges almost surely to $M$, as $n\rightarrow \infty$.

\section{Posterior Carachterization of a Norm-IFPP Process}
\label{sec:postnorm}
In this section we characterise the posterior distribution of the process $P\sim  \text{Norm-IFPP}(h,p_0, q_M)$. To this end, 
we introduce the random variable $U_n=\Gamma_n/T$, where $\Gamma_n\sim \text{Gamma}(n,1)$, with  $\Gamma_n$ and $T$ independent,  where  $T=\sum_{i \in \mathcal{M}} S_i $. 
It is easy to show (see the Appendix~\ref{sec:proof_marg_u}) that if $P\sim  \text{Norm-IFPP}(h,p_0, q_M)$ then, for any $n \ge 1$, the marginal density of $U_n$ is given by
\begin{equation}
  f_{U_n}(u;n) = \frac{u^{n-1}}{\Gamma(n)}(-1)^n \frac{d}{du^n}\E\left( 
\psi(u)^M \right)\\
\label{eq:margU}
\end{equation}
where $\psi(u)$ is the Laplace transform of the density $h$, as defined in Eq.~\eqref{eq:psi}. We give the derivation of Eq.~\eqref{eq:margU} in Appendix~\ref{sec:proof_marg_u}. The posterior distribution of $U_n$, given $\bm \theta =(\theta_1,\ldots,\theta_n)$, is crucial to perform posterior inference and allows us to derive the posterior distribution of the unnormalised process $\widetilde{P}$. To this end, we  need to show that \textit{a posteriori}, conditionally to $U_n$,  $\widetilde{P}$ is the superposition (union) of two  independent process: a point process and a finite process with fixed locations at $(\theta^\star_1,\ldots, \theta^\star_k)$. Note that $k$ corresponds to the number of allocated jumps $M^{(a)}$ and $M$ is equal to the sum of $k$ and the number $M^{(na)}$ of unallocated jumps, assuming values in $\mathbb{N}\cup \{0\}$. The process of unallocated jumps is a latent variable which links the parametric part of the model in $P$ to a nonparametric process. This link is essential for computations as it will become clearer in Section \ref{sec:posterior_inference}, where we discuss the algorithm. The results below are conditional on the realizations of the random variable $U_n$, which is a typical strategy in the theory of normalised random measures, since working on the augmented space allows us to exploit the quasi-conjugacy of the process $P$ \citep[see][]{james2009posterior}. We now present the main theoretical contribution of this work. 
\begin{theorem}
  \label{theo:Posterior}
  If $P\sim  \text{Norm-IFPP}(h,p_0, q_M)$, then the 
  unnormalized  process $\widetilde P$,  given
  ${\bm\theta}^\star=(\theta^\star_1,\dots,\theta^\star_k)$, $\bm n=(n_1,\dots,n_k)$ 
  and $U_n=u$, 
  is the superposition of two processes:
  $$  \widetilde P \stackrel{d}=
    \widetilde P^{(na)}\cup \widetilde P^{(a)}
   $$
  where 
  \begin{enumerate}
    \item \label{prop:noalloc} The process of non-allocated jumps $\widetilde P^{(na)}$  is an independent finite point process with Janossy density given
      by 
      \begin{equation*}
	j_m((s_1,\tau_1),\dots,(s_m,\tau_m)) =
	m!q_m^{\star}\prod_{j=1}^{m}h^{\star}(s_j)p_0(\tau_j)
      \end{equation*}
      where  
	$h^{\star}_{u}(s)\propto \e^{-us}h(s)$, $q^{\star}_m\propto \frac{(m+k)!}{m!} \psi(u)^{m}q_{m+k}$, $\psi(u)$ is the Laplace transform of $h$, and $m$ is a realization of $M^{(na)}$, the number of unallocated jumps, taking values in   $\{0,1,2,\dots\} $.
    \item  \label{prop:alloc} 
      The process of allocated jumps  $\widetilde P^{(a)}$ is the unordered set of points  $(S_1,\tau_1),\dots,(S_k,\tau_k)$, such that, for $j=1,\dots,k$,
      $\tau_j=\theta^{\star}_j$ and  the distribution of $S_j$ is proportional to
      $s^{n_j}\e^{-u s}h(s)$.
    \item Conditionally on $\mathcal M^{(a)}$ and $U_n=u$, $\widetilde P^{(a)}$ and $\widetilde P^{(na)}$ are independent.
  \end{enumerate}  
    \noindent Moreover,  the posterior law of $U_n$ given  ${\bm\theta}=(\theta_1,\ldots, \theta_n)$ depends only on the partition $\rho_n$ and  has density on the positive 
      reals given by
      \begin{equation*}
	f_{U}(u\mid  \rho_n )\propto 
	\dfrac{u^{n-1}}{\Gamma(n)}
	\left\{ 
	  \sum_{m=0}^{\infty}\frac{(m+k)!}{m!}\psi(u)^{m}q_{m+k}
	\right\}
	\prod_{i=1}^k 
	\kappa(n_j,u)
	\label{prop:margU}
      \end{equation*}
\end{theorem} 
\noindent \textit{Proof:} See Appendix~\ref{sec:app_proof_post_char} $\blacksquare$
  
The result in Theorem \ref{theo:Posterior} is the finite dimensional counterpart of Theorem 1 in \cite{james2009posterior} for normalised completely random measure. This theorem will allow building an efficient block Gibbs sampler for finite mixture models. 
Since the order in which the points of a point process arise is not important,  without loss of generality, given a realization
of the posterior process $\widetilde P$, we assume that,  in 
$\widetilde P=\{S_1,\dots,S_M\}$, $M=k+M^{(na)}$, i.e. the first $k$ points $\{S_1,\dots,S_k\}$ correspond to the allocated jumps, while the last $M^{(na)}$ to the non-allocated ones.

\section{Posterior inference}
\label{sec:posterior_inference}
To perform posterior inference tailored MCMC algorithms need to be devised. The two most popular strategies in Bayesian nonparametrics are marginal \citep{neal2000markov} and conditional algorithms \citep{ishwaran2001gibbs,kalli2011slice,argiento2016}. Our construction allows for straightforward extension of such startegies to the finite mixture case, offering a convenient alternative to the often inefficient and labour intensive reversible jump.  To implement marginal algorithms it is desirable (although not necessary, but at the cost of extra computations) to be able to compute the sum in Eq.~\eqref{eq:eppf_expr} to obtain the probability of a random partition. On the other hand, for conditional algorithms we need to  sample from the posterior distribution of a Norm-IFPP which requires  a closed form expression for the Laplace transform in Theorem \ref{theo:Posterior}. More specifically, it is essential to be able to sample from the posterior distribution of the number of the non-allocated jumps, $q^\star_m$, as well as from the distribution of the allocated and unallocated jumps, i.e. the  densities proportional to $e^{-us} h(s)$ (Exponential tilted) and $s^{n_j} e^{-us} h(s)$ (Gamma tilted). Specific solutions for well known processes  will be presented  in the following sections. Here we give a general outline of both algorithms.

\subsection{Marginal Algorithm}
\label{subsec:margalg}
As  mentioned before,  a sample $\theta_1,\dots,\theta_n$ from $P$ induces a partition of the set of the data indexes, denoted by $\rho_n= \{C_1,\dots, C_k\}$, such that $i\in C_j$ implies that datum $i$ belongs to cluster $j$. Marginal algorithms rely on the fact that, by integrating out the measure $P$, the only parameters left in Eq.~\eqref{eq:modello} are the random partition $\rho_n$ and the cluster specific parameters $\theta_1^\star,\dots,\theta_k^\star$.
Posterior sampling strategies for $\rho_n$ are based on the Chinese restaurant process \citep{aldous1985exchangeability}, which describes the (a priori) predictive generative process for $\rho_n$, and relies on the evaluation of the eppf associated with  $P$. Nevertheless, when $P$ corresponds to the Norm-IFPP model, this evaluation can be computationally burdensome due to the integral with respect to $u$ in Eq.~\eqref{eq:eppf_expr}. To design efficient algorithms we adopt a \emph{disintegration} technique following a strategy similar to the one suggested by \cite{james2009posterior} and \cite{favaro2013mcmc} for NRMI. In particular, we augment the state space introducing  the latent variable $U$ (see Theorem~\ref{theo:Posterior}). 

We now explain how, conditional to the latent variable $U_n:=U$, the Chinese restaurant process can be adapted to this set-up. Recall that the marginal distribution of $U_n$, defined in Section \ref{sec:nfpd}, with  $U_n\mid T\sim \text{Gamma}(n,T)$, has been derived in Eq.~\eqref{eq:margU}: $  f_{U_n}(u;n) = \frac{u^{n-1}}{\Gamma(n)}(-1)^n \frac{d}{du^n}\E\left(\psi(u)^M \right)$. 

The partition (or clustering) $\rho_n$ can be generated using the eppf derived in Theorem~\ref{thm:eppf_peps}. It is straightforward to show that
\begin{equation}
\begin{split}
  \calL(\rho_n,U_n)= \pi(n_1,\dots,n_k;u)  &=  
    \dfrac{u^{n-1}}{\Gamma(n)}
    \left\{ 
      \sum_{m=0}^{\infty}\frac{(m+k)!}{m!}\psi(u)^{m}q_{m+k}
    \right\}
    \prod_{j=1}^k 
    \kappa(n_j,u)  \\
    & =   \dfrac{u^{n-1}}{\Gamma(n)} \Psi(u,k) \prod_{j=1}^k 
    \kappa(n_j,u) 
\end{split}
    \label{eq:eppf_disintegrata}
\end{equation}
where $\Psi(u,k):=   \left\{ 
      \sum_{m=0}^{\infty}\frac{(m+k)!}{m!}\psi(u)^{m}q_{m+k}
    \right\}$.
 This joint distribution allows us to derive the predictive probability (conditionally on $U_n=u$) that observation $n+1$ belongs to a new cluster $C_{k+1}$ is  
\begin{align}
  \Prob(n+1\in C_{k+1} |u,\rho_n)&\propto
  \frac{\pi(n_1,\dots,n_k,1;u)}{\pi(n_1,\dots,n_k;u)}=\frac{S(u,k+1)}{S(u,k)}\kappa(1,u)
  \label{eq:newchin}
\end{align}
while the predictive probability of belonging to an existing cluster is
\begin{align}
  \Prob(n+1\in C_j |u,\rho_n)&\propto
  \frac{\pi(n_1,\dots,n_j+1,\dots,n_k,1;u)}{\pi(n_1,\dots,n_j,\dots,n_k;u)}=
  \frac{\kappa(n_j+1,u)}{\kappa(n_j,u)}, \quad j=1,\dots,k
\label{eq:oldchin}
\end{align}

As in a standard Chinese restaurant process \citep{aldous1985exchangeability}, a sequence of customers (data $i=1,2,\dots$) enter a restaurant
with an infinite number of  tables (groups $C_1, C_2,\dots$). The first customer sits at the first table and a random variable $U_1$ is drawn. Then  each subsequent customer joins a new table with probability proportional to Eq.~\eqref{eq:newchin}, or an existing table with $n_j$  customers with probability proportional to Eq.~\eqref{eq:oldchin}. For each new customer $i$, a variable $U_i$ is drawn. After $n$ customers have  entered the restaurant, the seating arrangement of customers around tables corresponds to a partition $\rho_n$ of $\{ 1, \ldots, n\}$ with numerosity $(n_1, \ldots, n_k)$, $n_j =\#C_j $.
The seating arrangement of the customers is exchangeable, in the sense that any seating that leads to the same number of occupied tables and the same number of customers per table has the same probability. The main difference with the standard Chinese process consists in updating the cluster allocation conditional on $U_i$'s. The strategy of conditioning on a sequence of auxiliary variables to generalise the Chinese restaurant process was introduced for infinite dimensional measures by \cite{james2009posterior}. Here, we have derived the finite dimensional counterpart.

A general scheme to implement a posterior Gibbs sampler for Norm-IFPP mixture model is the following:
\begin{itemize}
  \item[i.]  Draw $\rho_n$ from $\mathcal{L}(\rho_n\mid \text{rest})$. This can be done, for instance, using one of the several algorithms presented in \cite{neal2000markov}, by simply substituting the predictive distributions of the Dirichlet process with the conditional predictive structure of a Norm-IFPP given in Eq.s~\eqref{eq:newchin} and \eqref{eq:oldchin}.
  \item[ii.] Draw $U_n$ from $\mathcal{L}(U_n\mid \text{rest})$. This update requires a Metropolis step (or any other alternative that ensures that the  chain is invariant) with target distribution  proportional to $ \calL(\rho_n,U_n)$  in Eq.~\eqref{eq:eppf_disintegrata}
  \item[iii.] Draw $\theta_j^\star$, for each $j=1,\dots,k$,  from  $\mathcal{L}(\theta^\star_j\mid \text{rest})$. In general, this is straightforward and involves a simple  parametric update from
    \begin{equation*}
      \prod_{i\in C_j}f(y_i \mid  \theta_j^\star)p_0(\theta_j^\star)
    \end{equation*}
\end{itemize}
Special cases in which the full conditional distributions of $U_n$ and $\rho$ have a simple expression  will be discussed later.

\subsection{Conditional Algorithm}
\label{sec:condalg}
Conditional algorithms are usually of wide applicability. The most famous example of this type of strategy is the one proposed by \cite{ishwaran2001gibbs}, which consists of a blocked Gibbs sampler based on the stick-breaking representation of a discrete random measure. Conditional algorithms allow us to draw from the joint distribution of $(M,\mathbf{\tau},\mathbf{S}, \mathbf{c})$ in Eq.~\eqref{eq:latvar}, where $w_i=S_i/ T$, which in turns defines a draw  of the random probability measure on $\Theta$: 
$$ P(d\theta)=\sum_{m=1}^M w_m \delta_{\tau_m}(d\theta)$$ 
As the algorithm samples from the posterior distribution of the random measure, we are able to perform full posterior inference, at least numerically,  on any functional of such distribution. These issues are discussed in detail in \cite{gelfand2002computational}.   
Moreover, it is simple to make inference on the hyper-parameters of the distributions of $M$ and  $\mathbf{S}$. An outline of the MCMC algorithm is given in Figure~\ref{fig:schema_gibbs}. The scheme follows directly from Theorem~\ref{theo:Posterior}, adapted to the mixture case. Note that in step 2 of the algorithm, the relabelling of the  mixture components is essential so that the non-empty components correspond to the first $k$ components.
\begin{figure}[h!]
 \begin{flushleft}
 \begin{tikzpicture}[scale=0.7,transform shape]
 \path \start{0}{\textbf{Repeat} for g in 1...\textbf{G}:};
  \path (p0.south)+(9.0,-1.0)
  \etape{2}{\textcolor{blue!90!black}{\textbf{1.}} Sample ${u}^{(g)}$ from a Gamma$(n, T)$};
  \path (p2.south)+(0.0,-1.5) 
  \etape{3}{\textcolor{blue!90!black}{\textbf{2.}}
  For i=1,..,n sample ${c_i}^{(g)}$ from a discrete distribution s.t.
  $$\mathbb{P}(c_i=j\mid \text{rest}) \propto S_j f(y_i\mid  \tau_j), \quad j=1,\dots,M$$
After resampling the vector $\mathbf{c}^{(g)}$, calculate the number $k^{(g)}$ of unique values of $\mathbf{c}^{(g)}$ and relabel the mixture components in a way that the first $k^{(g)}$ ones are allocated. 		 		 
		 };
  \path (p3.south)+(0.0,-2.0)
  \etape{4}{\textcolor{blue!90!black}{\textbf{3.a}}
  Sample the  hyperparametrs $\eta_1^{(g)}$  of  the density $h$ from
  $$ \Prob(\eta_1=d\eta_1 \mid \text{rest}) \propto \left\{ 
	  \sum_{m=0}^{\infty}\frac{(m+k)!}{m!}\psi(u)^{m}q_{m+k}
	\right\}
	\prod_{j=1}^k 
	\kappa(n_j,u) \pi_1(\eta_1) d\eta_1 
   $$
   where $\pi_1(\eta_1)$ denotes the prior density for $\eta_1$.  
};

  \path (p4.south)+(0.0,-2.5)
  \etape{5}{\textcolor{blue!90!black}{\textbf{3.b}}
  Sample the  hyperparametrs $\eta_2^{(g)}$  of  the density $q_M$ from 
  $$ \Prob(\eta_2=d\eta_2 \mid \text{rest}) \propto \left\{ 
	  \sum_{m=0}^{\infty}\frac{(m+k)!}{m!}\psi(u)^{m}q_{m+k}
	\right\}
	\prod_{j=1}^k 
	\kappa(n_j,u) \pi_2(\eta_2)  d\eta_2
   $$
   where $\pi_2(\eta_2)$ denotes the prior density for $\eta_2$.  
};

  \path (p5.south)+(0.0,-1.2)
  \etape{6}{\textcolor{blue!90!black}{\textbf{4.a}}
  Sample $M^{(na)(g)}$ from 
   $$q^{\star}_m\propto \frac{(m+k)!}{m!} \psi(u)^{m}q_{m+k}$$
    and set $M^{(g)}=k^{(g)}+M^{(na)(g)}$
};
 
\path (p6.south)+(-6,-1.2)
  \decision{7}{\textcolor{blue!90!black}{\textbf{4.b.$i$}} 

  \textbf{Allocated jumps}: for $m=1,\ldots, k^{(g)}$, sample $S_m^{(g)}$ independently from \\ 
  \begin{center}
    $ \Prob(S_{m}=ds\mid \text{rest})  \propto$ $s^{n_m}\e^{-s u} h(s)ds$
  \end{center}
};
  \path (p6.south)+(6,-1.2)
  \decision{8}{\textcolor{blue!90!black}{\textbf{4.b.$ii$ }}
  \textbf{Non-allocated jumps}: for $m=k^{(g)}+1,\ldots, M^{(g)}$, sample $S_m^{(g)}$ independently from 
  \begin{center}
    $\Prob(S_m=ds\mid \text{rest}) \propto e^{-u s}h(s)ds$
  \end{center}
};
 
 \path (p7.south)+(0,-1.2)
 \decision{9}{\textcolor{blue!90!black}{\textbf{4.c.$.i$ }} 
   \textbf{Allocated points of support}:
   sample $\tau_m^{(g)}$ independently from 
   \begin{center}
   $\Prob(\tau_m=d\tau_m\mid \text{rest}) \propto 
   \left\{  \prod_{i \in C_m} f(y_i\mid \tau_m)   \right\} p_0(\tau_m)d\tau_m$ 
  \end{center}
 };
  \path (p8.south)+(0,-1.2)  
  \decision{10}{\textcolor{blue!90!black}{\textbf{4.c$.ii$ }}
  \textbf{Non-allocated points of support}: sample $\tau_m^{(g)}$ independently from the prior, i.e. 
 \begin{center}
   $\Prob(\tau_m=d\tau_m\mid \text{rest}) = p_0(\tau_m)d\tau_m$ 
  \end{center}
  }  ;
\end{tikzpicture}
\end{flushleft}
  \caption{Blocked Gibbs sampler scheme; the conditioning arguments of all full conditionals have been omitted to simplify notation.}
  \label{fig:schema_gibbs}
\end{figure}

\section{Norm-IFFP hierarchical mixture models}
\label{sec:nfpd}
Most real world applications of discrete random measures involve an additional layer in the model hierarchy and convolve the random measure with a  continuous kernel leading to nonparametric mixture models. In this context,  data are assumed to be generated from a parametric distribution indexed by some parameter $\theta$, with $\theta\sim P$. Usually $P$ is assigned a nonparametric prior, in our case  a Norm-IFFP. This leads to models of the form
\begin{equation}
  \begin{split}
    &Y_1,\dots,Y_n |\theta_1,\dots,\theta_n \stackrel{ind}{\sim}
    f(y \mid \theta_i)\\
    &\theta_1,\dots,\theta_n |P \stackrel{iid}{\sim}P\\
    &P\sim Norm-IFFP(h,p_0, q_m) 
  \end{split}
  \label{eq:NFDPMIX}
\end{equation}
where $f(\cdot \mid \theta_i)$ is a parametric  density on $\mathcal{Y}$, for all $\theta\in\Theta \subset \Rea^d$. We point out that $p_0$ is the density  of a non-atomic probability measure $P_0$ on $\Theta$, such that $\mathbb{E}(P(A))=P_0(A)$ for all $A\in\calB(\Theta)$. Model \eqref{eq:NFDPMIX} will be addressed here as a \textit{Norm-IFFP hierarchical mixture} model. The model can be extended by specifying appropriate hyperpriors. It is well known that this model is equivalent to assuming that the $Y_i$'s, conditional on $P$, are independently distributed according to the random density \eqref{eq:mixture1}. 
We point out that  Model \eqref{eq:NFDPMIX} admits as a special case  the popular finite Dirichlet mixture model (see \cite{nobile94,richgreen97,steph00,miller2017mixture}) discussed in more details in Section~\ref{sec:FDP}.
The posterior characterization given in Theorem~\ref{theo:Posterior}, as well as  the analytical expression for the eppf given in Theorem~\ref{thm:eppf_peps}, allow us  device  conditional or marginal algorithms to perform inference under Model~\eqref{eq:NFDPMIX} as discussed in Section~\ref{sec:posterior_inference}.

\section{Special Choices of $q$ in Norm-IFFP} 
\label{sec:special_q}
The exact evaluation of the eppf in Eq.~\eqref{eq:eppf_expr} presents two challenges:  an integral and an infinite sum. Numerical solution of  the integral is handled within the MCMC via the augmentation trick, while here we discuss more in detail the infinite sum, defined as 
\begin{equation*}
  \Psi(u,k):=
 \left\{ 
      \sum_{m=0}^{\infty}\frac{(m+k)!}{m!}\psi(u)^{m}q_{m+k}
    \right\}
\end{equation*}
for each real $u>0$ and each integer $k\ge1$.  As it is shown in the proof of Theorem~\ref{theo:Posterior}, $\Psi(u,k)\le\frac{k!}{(1-\psi(u))^{k+1}}$, i.e. the sum always converges.

The analytical solution of the latter depends on the particular choice of prior distribution $q$ for $M$. Since  $\psi(u)$ is less than 1, $ \Psi(u,k)$  is related to a binomial series. This implies that if $q_m$ is a  Poisson or a Negative Binomial, we can derive conjugate updates for $M^{(na)}$ and  we can find a closed form solution for $\Psi(u,k)$.
In particular,  if $q_{m}=\calP_1(m,\Lambda)$, corresponding to the density of  a random variable shifted on $\{1,2,\dots,\}$, then  we obtain
\begin{equation*}
   \Psi(u,k)=\Lambda^{k-1}(\Lambda \psi(u)+k)\exp\{\Lambda(\psi(u)-1)\}
\end{equation*}
Moreover, the full conditional distribution of $M^{(na)}$, i.e.  $q^\star_m$ in item (a) of Theorem~\ref{theo:Posterior}, is
\begin{equation*}
  q^\star_m\propto k \psi(u) \mathcal{P}_0(m,\Lambda \psi(u)) + 
	    \Lambda \psi(u) \mathcal{P}_1(m,\Lambda \psi(u) )
\end{equation*}
where $\mathcal{P}_0$ and $\mathcal{P}_1$ are the probability mass function of a Poisson and of a shifted Poisson respectively.

Finally, it is worth to mention that the shifted Poisson choice for $M$ implies that in Eq.~\eqref{eq:margU}, we have
\begin{align*}
  f_{U_n}(u;n)=
\frac{u^{n-1}}{\Gamma(n)}(-1)^n \frac{d}{du^n}
\psi(u)\,{{\e}^{\Lambda\, \left( \psi(u)-1 \right) }}
\end{align*}
Note that it is also possible to use a Truncated Poisson distribution for $M$, with a slight difference in results.

On the other hand if  we choose $q_{m}=\text{NegBin}(m;p,r)$,  a
Negative Binomial density with parameters $0\leq p\leq 1$ and $r>0 $ and support on $\{1,2,\dots\}$, i.e.
\begin{equation*}
  q_m=\frac{\Gamma(r+m-1)}{(m-1)!\Gamma(r)} p^{m-1}(1-p)^r\uno_{\{1,2,\dots\}}(m)
\end{equation*}
then, it is easy to show that 
\begin{equation*}
   \Psi(u,k)=\frac{\Gamma\left( r+k-1 \right)}{\Gamma  \left( r \right) }
  {p}^{k-1} \left( 1-p \right) ^{r}
  \frac{  p\psi(u)(r-1)+k }{
    \left(1 -p\,\psi(u) \right)^{k+r} }
\quad u>0 \text{ and }\quad k\ge1
\end{equation*}
In this case we obtain that the full conditional for the number of non-allocated components has support in $m=0,1,\dots$ with probability 
mass function
\begin{equation*}
  q^\star_m \propto (r+k)p\,\psi(u) \text{NegBin}(m;p\,\psi(u),r+k)+  k\left( 1-p\,\psi(u) \right)\text{NegBin}(m+1;p\,\psi(u),r+k-1)
\end{equation*}
Moreover,
\begin{align*}
  f_{U_n}(u;n)&=
\frac{u^{n-1}}{\Gamma(n)}(-1)^n \frac{d}{du^n}
{\frac {\psi(u)\, \left( 1-p \right) ^{r}}{ \left( 1-p\,\psi(u) \right) ^{r
}}} \\
\end{align*}

Finally, in applications,  we might want to fix the number of mixture components, i.e. the number of points of the point process,    leading to the  standard finite mixture setup. In this case, if  $M$ is set very large,  we recover the sparse mixture framework of \cite{fruhwirth2017here}.
Let $M= \widetilde{M} \ge1$ with probability 1, we obtain   
\begin{equation*}
  \Psi(u,k)=\left\{
  \begin{array}[h]{ll}
    \frac{\widetilde{M} !}{(\widetilde M-k)!}\psi(u)^{\widetilde{M} -k} & \text{if } k\le \widetilde{M} \\
    0 & \text{if } k>\widetilde{M} 
  \end{array}
\right.
\end{equation*}
This prior specification for $M$ implies that the support for $k$ is bounded,  $k\le \widetilde{M} $, $q^\star_m$ assigns probability mass one to $\widetilde M-k$  and 
\begin{align*}
  f_{U_n}(u)=
  \frac{u^{n-1}}{\Gamma(n)}(-1)^n \widetilde{M} \psi(u)^{\widetilde{M} -1}\psi'(u)
\end{align*}

\section{Important Examples}
\label{sec:impex} 
The ${\rm Norm-IFPP}(h,p_0,q_M)$ depends on three densities.  The prior on  $\theta_i$ is $p_0$  and, in applications, a conjugate prior is usually preferred. The choice of $q_M$ has been discussed in Section \ref{sec:special_q}.  We now focus on the choice of $h$. Once again the particular choice of $h$ influences the induced clustering in Eq.~\eqref{eq:eppf_expr} as well as efficiency of computations. There are two possible alternatives: either to choose as $h$  a parametric density or to select the Laplace transform of $h$, $\psi(u)$. 

\subsection{Choice of $h$}
\subsubsection{Finite Dirichlet process}
\label{sec:FDP}
Let $h$  be the  $\text{Gamma}(\gamma,1)$  density. Under this choice of $h$ the Norm-IFPP is a finite Dirichlet measure, that is an almost surely discrete probability measure as in Eq.~\eqref{eq:nor_ifpp}, where, conditionally on $M>0$,  the jump sizes  $(w_1,\dots,w_M)$ of  $P$  are a sample from the $M$-dimensional Dirichlet$_{M}(\gamma,\dots,\gamma)$ distribution. Therefore, this is equivalent to a conventional finite mixture model as described in Section \ref{sec:fmm}. Recall that the Laplace transform and its derivatives for a $\text{Gamma}(\gamma,1)$ density are given by
$
\psi(u)=\frac{1}{(u+1)^\gamma},\ \text{and}\ 
  \kappa(n_j,u)=\frac{1}{(u+1)^{n_j+\gamma}}\frac{\Gamma(\gamma+n_j)}{\Gamma(\gamma)},\quad
  u>0,\  n_j=1,2,\dots
$
Then,  applying Theorem~\ref{thm:eppf_peps}, we obtain that the
eppf of this model is
\begin{eqnarray}
\nonumber
  p(n_1,\dots,n_k)&=& \int_{0}^{\infty} \frac{u^{n-1}}{\Gamma(n)} \left(
  \sum_{m=0}^{\infty}\frac{(m+k)!}{m!}\frac{1}{(u+1)^{m\gamma}} q_{m+k}
  \right)\prod_{j=1}^{k}\frac{\Gamma(\gamma+n_j)}{\Gamma(\gamma)}\frac{1}{(u+1)^{n_j+\gamma}}du\\
  &=&
\nonumber
\left\{  
  \frac{1}{\Gamma(n)}\sum_{m=0}^{\infty}
  \frac{(m+k)!}{m!}q_{m+k}\int_{0}^{\infty}\frac{u^{n-1}}{(u+1)^{m\gamma+n+kn}}du 
  \right\}
  \prod_{j=1}^{k}\frac{\Gamma(\gamma+n_j)}{\Gamma(\gamma)}\\
  &=& 
 \left\{ 
  \sum_{m=0}^{\infty} 
  \frac{(m+k)!}{m!}q_{m+k} \frac{\Gamma((k+m)\gamma )}{\Gamma( (k+m)\gamma+n)} 
   \right\}
  \prod_{j}^{k}\frac{\Gamma(\gamma+n_j)}{\Gamma(\gamma)}
  \label{eq:eppf_dir}
  \\
 \nonumber
  &=&
V(n,k)\prod_{j=1}^{k}\frac{\Gamma(\gamma+n_j)}{\Gamma(\gamma)}
\end{eqnarray}
See also Chapter 2 in \cite{pitman2006combinatorial} and \cite{miller2017mixture}. 
  In particular, if we choose \emph{semi-conjugate} priors for the number of components as we discussed in 
  Section~\ref{sec:special_q},
we can obtain integral representations for $V(n,k)$. When $q_m$ is shifted Poisson distribution with parameter $\Lambda$,
$$  V(n,k)=\Lambda^{k-1}\int_{0}^{\infty}\frac{u^{n-1}}{\Gamma(n)}\frac{\Lambda+k(u+1)^{\gamma}}{(u+1)^{n+\gamma(k+1)}}
\exp\left\{-\Lambda\frac{(u+1)^{\gamma}-1}{(u+1)^{\gamma}}  \right\}du
$$
while when  $q_m$ is a Negative Binomial with with parameters $p$ and $r$ 
$$  V(n,k)=
\frac{\Gamma(r+k-1)}{\Gamma(r)} {p^{k-1} (1-p)^r}  \int_{0}^{\infty}\frac{u^{n-1}}{\Gamma(n)}\frac{p(r-1)+k(u+1)^{\gamma}}{(u+1)^{n+\gamma(r-1)}}\frac{1}{ 
\left( (u+1)^\gamma-p \right)^{r+k}}du
$$
Finally when $q_m$ assigns probability one to $\widetilde M$
$$V(n,k)=
\frac{\widetilde M!}{(\widetilde M-k)!}\frac{\Gamma(\gamma\widetilde M)}{\Gamma(n+\gamma\widetilde M)}\uno_{\{1,\dots,\widetilde{M}\}}(k)
$$
The finite Dirichlet process case has been extensively discussed by \cite{miller2017mixture}. They propose a marginal algorithm which  requires  evaluating the sum in Eq.~\eqref{eq:eppf_dir}. This restriction implies that  an approximation of the infinite sum needs to be evaluated at every step of the algorithm, slowing down computations for large $n$ and making it difficult to specify a prior on $\gamma$ and on the number of allocated components.  We note that these difficulties are easily overcome, for convenient  choices of $q_m$, by employing the disintegration trick and implementing a conditional MCMC scheme, as described in Section~\ref{sec:condalg}.
We highlight that when $q_m$ is a point mass, the marginal algorithm becomes even more straightforward as we can obtain a closed form expression for $V(n,k)$. 

Note that, when $q_m$ is a shifted Poisson, if $\gamma=\alpha/\Lambda$,  for $\alpha>0$, and $\Lambda\rightarrow \infty$, then $P$ converges in distribution to the Dirichlet process with  mass parameter $\alpha$ (see Appendix~\ref{app:conv} for a proof). Similarly, we recover the Dirichlet process when $q_m$ assigns mass one to $\widetilde{M}$, $\gamma=\alpha/\widetilde{M}$ and $\widetilde{M}$ goes infinity. This case has been extensively investigated in the Bayesian nonparametric literature from both computational and methodological perspective \citep[see][for a thourough discussion]{ishwaran2002exact}.

Furthermore, Eq.~\eqref{eq:eppf_dir} implies that the finite Dirichlet process is a member of the family of Gibbs partition distributions \citep{pitman96,lijoi2010models}. 
The Gibbs type structure allows us to simplify the prior for  number of occupied components given in Eq.~\eqref{eq:priorK}, which in the  FDMM case becomes
\begin{equation}
\label{eq:priorK}
p^\star_k=\Pr\{M^{(a)}=k\}=
V(n,k)\gamma^k S_{n,k}^{-1,\gamma}    
\end{equation} 
 where $S_{n,k}^{-1,\gamma}$ is the \emph{Generalized Stirling number} computed for $k$ compositions of $n$ with parameters $-1$ and $\gamma$, i.e.
 \begin{equation*}
   S_{n,k}^{-1,\gamma}=\sum_{j=k}^{n}(-1)^{n-j}s_{n,j}S_{j,k}\gamma^{j-k}
 \end{equation*}
 where $s_{n,j}$ is a Stirling number of the first kind and $S_{j,k}$ is a Stirling number of second kind as defined in Eq. (1.16) and (1.13) of \cite{pitman2006combinatorial} respectively.

Finally, since the finite Dirichlet process is widely used in applications, we give in Section~\ref{sec:alg_fmm} of the Appendix a detailed description of the conditional algorithm  when $q_M$ is the density of a shifted Poisson distribution and appropriate hyperpriors are specified on  $\gamma$ and $\Lambda$.

\subsubsection{Uniform weights} 
Let $$h(s)=\uno_{(0,1)}(s)$$ i.e. the un-normalized jumps are uniformly distributed. To implement the conditional algorithm and compute the eppf  we need to evaluate the Laplace transform as well as its derivatives of degree $n$ for each $n>1$. To this end, we need to solve, for each $n\ge0$, the following integral
 \begin{align}
   \int_{0}^{\infty}s^{n}\e^{-us}h(s)ds=\int_{0}^{1}s^n\e^{-us}ds=
   \frac{1}{u^{n+1}}\int_{0}^{u}z^{n}e^{-z}dz=\frac{\gamma(n+1,u)}{u^{n+1}}
 \label{eq:unifweight}
 \end{align}
where $\gamma(\alpha,u)=\int_{0}^{u}z^{\alpha-1}\e^{-z} dz$ is the upper-incomplete gamma function \citep{GraRyz07}. Moreover,  for $n\ge0$, the upper incomplete gamma function simplifies to
 \begin{equation*}
   \gamma(n+1,u)=n!\left[ 1-\e^{-u}\sum_{m=0}^{n}\frac{u^m}{m!} \right]
 \end{equation*}
 Exploiting the above result leads to
 \begin{equation*}
   \begin{array}[h]{ccc}
     \psi(u)=\frac{1-e^u}{u}, & \text{and} & \kappa(n_j,u)=\frac{\gamma(n_j+1,u)}{u^{n_j+1}}
   \end{array}
 \end{equation*}
 The evaluation of $\psi$  is essential to implement the conditional alogorithm in the case of uniform weights. Moreover, an efficient implementation of the conditional algorithm requires us to be able to sample from  the tilted Gamma version of $h$ in Eq.~\eqref{eq:unifweight}, which in this case is simply a truncated Gamma distribution on $(0,1)$.

 Applying Eq.~\eqref{eq:eppf_expr},  we can also easily obtain the eppf of the process with uniform jumps 
 \begin{equation*}
   \pi(n_1,\dots,n_k)=\int_{0}^{\infty}\frac{u^{n-1}}{\Gamma(n)}\frac{\Psi(n,k)}{
     u^{n+k}}\prod_{j=1}^{k}\gamma(n_j+1,u)
 \end{equation*}
 where $\Psi(n,k)$ is defined in Section~\ref{subsec:margalg}.

\subsubsection{Gamma approximation}
Any absolutely continuous  density on $\Rea^+$ can be approximated by a mixture of Gamma densities. Indeed,
\citet[p.14]{devore1993constructive}  showed that if $h(s)$ defined on $(0,\infty)$ has limit zero as
 $s\rightarrow \infty$, then $\mathscr{S}_\varepsilon(s)$, defined as 
    \begin{equation}
      \label{eq:gamma_apprx_h}
      \mathscr{S}_\varepsilon(s)=\e^{-s/\varepsilon}\sum_{l=0}^{\infty}\frac{s^l}{\varepsilon^l l!}h\left(\varepsilon l \right) 
      =\sum_{l=0}^{\infty}\varepsilon h(\varepsilon l) \text{Gamma}(s;l+1,\frac{1}{\varepsilon})
    \end{equation}
admits as limit
 \begin{equation*}
   \lim_{\varepsilon\rightarrow 0}\mathscr{S}_\varepsilon(s)=h(s)
  \end{equation*}
  uniformly for $0 < s < \infty$.  
Therefore, $h$ can be approximated by a mixture of Gamma densities, i.e Gamma$(s; l + 1, 1/\varepsilon)$, with unnormalized weights
 $\varepsilon h(\varepsilon l)$, for $l=0,1,\dots$ . 
 See \cite{wiper2001mixtures} for an extensive discussion of mixtures of Gamma  distributions and their convergence properties.  This is a powerful result as we can approximate any $h$ with a mixture of Gamma densities and allows us to consider a large class of weight distributions at the cost of computational complexity. In practice,  to approximate $h$ we need to set a tolerance level $\varepsilon$ sufficiently small. Let  
  $\mathscr{C} = \left(  \sum_{l=0}^{\infty}\varepsilon h(\varepsilon l)\right)^{-1}$. It is easy to show that
 \begin{equation*}
   \psi(u) = \mathscr{C} \sum_{l=0}^{\infty}\left( \frac{1}{1+\varepsilon u}\right)^{l+1}\varepsilon h(\varepsilon l)
 \end{equation*}
Moreover, it is obvious that $\e^{-us}\mathscr{S}_\varepsilon(s)$ and 
 $s^{n_j}\e^{-us}\mathscr{S}_{\varepsilon}(s)$  are both infinite mixtures of Gamma densities. The normalising constant of the first one is $\psi(u)$, while the normalizing constant of the second one is given by the function $\kappa$:
 \begin{equation*}
   \kappa(n_j,u)=
   \mathscr{C} 
  \left(   
   \frac{\varepsilon}{1+u\varepsilon}
   \right)^{n_j}
   \sum_{l=0}^{\infty}
   \frac{\Gamma(n_j+l+1)}{\Gamma(l+1)}
   \left( \frac{1}{1+\varepsilon u}
   \right)^{l+1}\varepsilon h(\varepsilon l)
 \end{equation*}

\subsection{Choice of $\psi$: Point processes with infinite divisible jumps}

It is well known \cite[it follows from the Levy-Khinchine formula in][]{jacod2013limit}, that the Laplace transform of an infinite divisible random variable $S$ has the form 
\begin{equation*}
  \psi(u)=\exp\left\{-\int_{0}^{\infty}
  (\e^{uz}-1) \omega(z)dz \right\}
\end{equation*}
where  the \emph{L\'evy intensity} $\omega(z)$  is the intensity of a measure on  the positive reals satisfying the regularity condition $\int_{0}^{\infty }\min(1,z)\omega(dz)<\infty$. Moreover, if  the distribution of $S$ is absolutely continuous (with respect to the Lebesgue measure) with  strictly positive  and continuous density $h(s)$ on $\Rea^+$, then $\int_{0}^{\infty}\omega(z)dz=\infty$. As an alternative to specifying $h$ we can choose $\omega$ that uniquely identifies $h$, for instance via the integral equation:
\begin{equation*}
  h(s)=\int_{0}^{s}\omega(z)h(s-z)\frac{z}{s}dz
\end{equation*}
The theory of positive infinite divisible random variables has been very useful for the study of Normalised Random Measures with Independent  Increments \citep{regazzini2003distributional}, because $S$ can be written as an infinite sum of positive random variables, i.e. $0<S=\sum_{j=1}^{\infty}S_j<\infty$. For the Norm-IPPF, we define the law of the unnormalised jumps $S_j$, which then   defines the distribution   of $S$. This is different from the work of \cite{regazzini2003distributional} where the distribution of the weights of the a.s. discrete random measure is derived by first specifying the law of the normalising constant $S$. In detail,  we want to assign the density $h$ of the unnormalized weights $S_j$ in Section \ref{sec:nifpp} such that the Laplace transform has a closed form and posterior inference is computationally manageable. We point out how, once we have a closed form of the Laplace transform $\psi(u)$ and of its derivatives $k(n_j,u)$, we can easily compute the eppf using Eq.~\eqref{eq:eppf_expr} so that marginal algorithms are straightforward to implement  as discussed in Section \ref{sec:posterior_inference}. On the other hand, if we are able to build a sampler to draw from  the tilted density $\e^{-us}h(s)$ and from the Gamma tilted density $s^{n_j}\e^{-us}h(s)$ the implementation of a conditional algorithm is straightforward. In the following we present three relevant examples for which computations are feasible.

\subsubsection{Gamma Process} 
For the particular choice of
\begin{eqnarray}
  \omega(z;\gamma)=\gamma z^{-1}e^{-z}
  \label{eq:gamma_levy_intensity}
\end{eqnarray}
where $\gamma > 0$, we obtain that the density $h$  of $S_j$ coincides with a Gamma density with parameters $(\gamma,1)$ density. This is the exact same situation of \ref{sec:FDP}.

\subsubsection{$\sigma$-Stable Process}
Consider the  Levy density 
\begin{equation}
  \omega(z;\sigma)=\frac{\sigma z^{-\sigma-1}}{\Gamma(1-\sigma)}
  \label{eq:stable_levy_intensity}
\end{equation}
with $0< \sigma <1$. It is straightforward to show that
\begin{align*}
  -\log(\psi(u))=\int_{0}^{\infty}(1-\e^{u z})\omega(z,\sigma)dz=u^{\sigma}
\end{align*}
and  the Laplace transform is
\begin{align}
  \psi(u)=\exp(-u^{\sigma})
  \label{eq:stable_laplace_transform}
\end{align}
\citet{pollard46} shows 
 that the density of $S_j$ with Laplace transform in Eq.~\eqref{eq:stable_laplace_transform} can be represented as follows:
\begin{align*}
  h(s;\sigma)=\frac{-1}{\pi}\sum_{k=0}^{\infty}\frac{(-1)^k}{k!}\sin(\pi\sigma
  k)\frac{\Gamma(\alpha k+1)}{s^{\sigma k +1}}
\end{align*}
Although the density $h$ is computationally intractable, since a closed form expression for the Laplace transform is available, it is possible to implement a marginal algorithm by calculating the derivatives of $\psi$. Exploiting Eq~~(13) in \cite{favaro2015random} we obtain that
\begin{equation*}
  k(n_j,u)=\frac{\e^{-u^{\sigma}}}{\sigma^{n_j}}\sum_{k=1}^{n_j}u^{\sigma k}\mathcal{C}(n_j,k;\alpha,0) 
\end{equation*}
where, for any non-negative integer $n\ge 0$,  $0\le k\le n$,  real numbers $\alpha$, $\beta$,  $\mathcal{C}(n,k;\alpha,\beta)$ denotes the noncentral generalized factorial coefficient (see \cite{charalambides2005combinatorial} for details). Here we mention that these indices can be easily computed when $\beta=0$ using the recursive formula
\begin{equation*}
  \mathcal{C}(n,k,\alpha,0)=\alpha \mathcal{C}(n-1,k-1;\alpha,0)
  +(n-1-k\alpha)\mathcal{C}(n-1,k;\alpha,0)   
\end{equation*}
with $\mathcal{C}(1,1,\alpha,0)=\alpha$. 

To implement a conditional algorithm we need to be able to sample from an Exponential  tilted stable density $\e^{-us}h(s)$ (also known as generalized Gamma) as well as from a Gamma tilted density $s^{-n_j}\e^{-us}h(s)$. Strategies to sample from an Exponential tilted density are presented in \cite{devroye2009random} and \cite{hofert2011sampling} while a method to sample  from the Gamma tilted $\sigma$-stable density is discussed in  Section 3.1 of \cite{favaro2015random}.

\subsubsection{Bessel Process}
Consider the intensity
\begin{equation*}
\label{eq:Bessel_intensity}
\omega(z; \alpha,\beta)= \frac{\alpha}{z}\e^{-\beta z} I_0(z), \quad z>0
\end{equation*}
where $\beta\geq 1$, $\alpha>0$ and let 
\begin{equation*}
I_\alpha(z)= \sum_{l=0}^{+\infty}\frac{(z/2)^{2l+\alpha}}{l!\Gamma(\alpha+l+1)}
\end{equation*}
be the modified Bessel function of order $\alpha\ge0$ \citep[see][Section
7.2.2]{erdelyi53}. Then, for $s>0$, 
\begin{equation}
\label{eq:2super}
\omega(z;\alpha,\beta)=  \frac{\alpha}{z}\e^{-\beta z} +
\sum_{l=1}^{+\infty}\frac{\alpha}{2^{2l}(l!)^2}z^{2l-1} \e^{-\beta z}
\end{equation}
so that $\omega$ is the sum of the L\'evy intensity of a Gamma process 
 and of the L\'evy intensities 
\begin{equation}
\label{eq:bessel_rhom}
\omega_l(z;\alpha,\beta)= \frac{\alpha}{2^{2l}(l!)^2}z^{2l-1} \e^{-\beta z}, \quad z>0, \qquad l=1,2,\ldots
\end{equation}
corresponding to finite activity Poisson processes \citep[see][]{argiento2016}. 

\begin{proposition}
When $\alpha>0$, we have

\begin{enumerate}
  \item[(a)]  
    the density $h$ corresponding to the L\'evy intensity $\omega(z;\alpha,\beta)$ is
    \begin{equation*}
    h(s)=\alpha \left( \beta+\sqrt{\beta^2-1} \right)^\alpha 
\dfrac{\e^{-\beta s}}{s}I_{\alpha}(s),\ \ s>0
    \end{equation*}
\item[(b)] the Laplace transform of $S_m \sim h(s)$ is 
\begin{equation*}
  \psi(u) = \left( \frac{\beta+\sqrt{\beta^2-1}} {\beta+u+\sqrt{(\beta+u)^2-1}} \right)^\alpha
\end{equation*}
\item[(c)]  the function $\kappa(n_j,u)$ has the following expression
\begin{equation*}
  \kappa(n_j,u)=
  \frac{\alpha \left(  \beta+\sqrt{\beta^2-1} \right)^\alpha}{2^{\alpha}(u+\beta)^{n_j+\alpha}} \frac{\Gamma(\alpha+n_j)}{\Gamma(\alpha+1)} \ff \left(\frac{n_j+\alpha}{2}, \frac{n_j+\alpha+1}{2};\alpha+1; \frac{1}{(u+\beta)^2}\right) 
\end{equation*}
where
$$ \ff(\alpha_1,\alpha_2;\gamma;z) :=  \sum_{l=0}^{\infty}
\frac{  \left( \alpha_1\right)_l
\left( \alpha_2 \right)_l
}{\left( \gamma\right)_l}
\frac{1}{l!} \left( z\right)^l, \quad\textrm{ with }  (\alpha)_l :=\frac{\Gamma(\alpha+l}{\Gamma(\alpha)} $$ is the hypergeometric series \citep[see][Equation (9.100)]{GraRyz07}. 
\end{enumerate}
\label{prop:bessel}
\end{proposition}
\noindent \textit{Proof:} See Appendix $\blacksquare$

Using Eq.~\eqref{eq:eppf_expr} we can derive the eppf for the Bessel case: 
 \begin{eqnarray*}
 \pi(n_1,\dots,n_k)&  =  &
    \int_{0}^{+\infty}
    \dfrac{u^{n-1}}{\Gamma(n)}
     \Psi(u,k)
    \prod_{j=1}^k 
    \kappa(n_j,u)du \\ &=& 
     \int_{0}^{+\infty}
    \dfrac{u^{n-1}}{\Gamma(n)(u+\beta)^{n+k\alpha}}
  \left(\frac{\alpha (\beta+\sqrt{\beta^2-1 } )^{\alpha } }{2^\alpha} \right)^k   \Psi(u,k) \\ &\cdot & 
    \prod_{j=1}^k \frac{\Gamma(\alpha+n_j)}{\Gamma(\alpha)}   \ff \left(\frac{n_j+\alpha}{2}, \frac{n_j+\alpha+1}{2};\alpha+1; \frac{1}{(u+\beta)^2}\right) d u 
\end{eqnarray*}
Therefore, the difficulty in implementing a marginal algorithm is simply in the evaluation of the $\ff$ function.

To implement a  conditional algorithm, also in this case we need to  
sample from an Exponential tilted density $\e^{-us}h(s)$ and a Gamma tilted
 $s^{n_j}\e^{-us}h(s)$ density for each real $u>0$ and integer $n_j>1$. From the proof of Proposition \ref{prop:bessel}, it is clear that
 these densities are both a mixture of Gamma distributions. In particular, we have that
 $$\e^{-us}h(s)\propto 
 \sum_{l=1}^{\infty}
 o_{l}(u)\text{Gamma}(s;2l+\alpha,u+\beta)
 $$
 and
 $$s^{n_j}\e^{-us}h(s)\propto \sum_{l=1}^{\infty}
 o_{l}(u,n_j)\text{Gamma}(s;2l+n_j+\alpha,u+\beta)
 $$
 where the mixture weights are given by 
 \begin{align*}
  &o_l(u)=\frac{\alpha}{2^{\alpha}}\left( \frac{\beta+u+\sqrt{(\beta+u)^2-1}}{\beta+u} \right)^\alpha\frac{1}{l!\Gamma(\alpha+l+1)2^{2l}}\frac{\Gamma(2l+\alpha)}{(\beta+u)^{2l}}\\
   &o_{l}(u,n_j)=\frac{1}{ \ff \left(\frac{n_j+\alpha}{2}, \frac{n_j+\alpha+1}{2};\alpha+1; \frac{1}{(u+\beta)^2}\right)}
\frac{\Gamma(\alpha+1)}{\Gamma(\alpha+n_j)}
\frac{1}{l!\Gamma(\alpha+l+1)2^{2l}}
\frac{\Gamma(2l+\alpha+n_j)}{\left( u+\beta \right)^{2l}}
 \end{align*}
 To sample from these two infinite mixtures we can use rejection samplings as the weights go quickly  to zero.  

\section{Galaxy data}
    \label{sec:Galaxy}
    We illustrate our model using the Galaxy dataset \citep{roeder1990density}, which offers a standard benchmark for mixture models. It contains $n=82$ measurements on  velocities of different galaxies from six well-separated conic sections of space. Values are expressed in Km/s, scaled by a factor of $10^{-3}$. We fit Model \eqref{eq:NFDPMIX}, using a Gaussian density $\calN(\mu,\sigma^2)$ on $\mathbb{R}$ as $f(y \mid \tau)$, $\tau=(\mu,\sigma^2)$. We specify the following prior  $p_0(\mu, \sigma^2) =\calN(\mu;m_0,\sigma^2/\kappa_0) \times \text{Inv-gamma}(\sigma^2; \nu_0/2,\nu_0/2\sigma^{2}_0)$.
    Here $\text{Inv-Gamma}(a,b)$ denotes the Inverse-Gamma distribution with mean $b/(a-1)$ (if $a>1$). We set $m_0=\bar x_n=20.8315$, $\kappa_0=0.01$, $\nu_0=4$, $\sigma_0^2=4$. Finally we assume a shifted $\text{Poisson}(\Lambda)$ as prior on $M$ and a Gamma$(\gamma,1)$ as a prior for $S_m$ (i.e. a finite Dirichlet process as mixing distribution). We implement the conditional algorithm described in Appendix~\ref{sec:alg_fmm} to perform posterior inference. In particular, we focus on  density estimation and inference on the number of mixture components and clusters. 
    
\begin{figure}[h!]
      \centering
     \includegraphics[width=\textwidth]{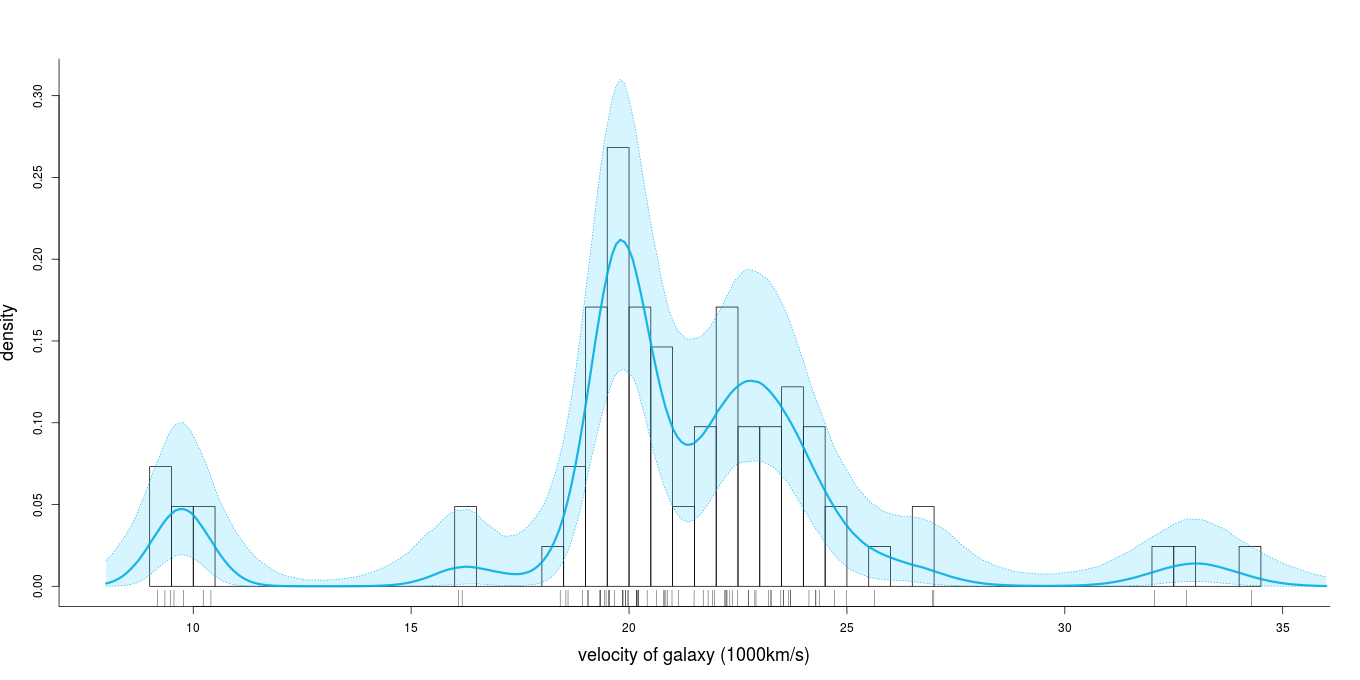}
      \caption{Density estimation  for $m_0=\bar x_n=20.8315$,
	$\kappa_0=0.01$, $\nu_0=4$, $\sigma_0^2=0.5$. The hyperparameter settings of the mixing distribution are specified in simulation scenario in $D.1$ corresponding  to the optimal value of the LPML index.}
      \label{fig:dens}
    \end{figure}

    First of all, we fit the model with $\Lambda$ and $\gamma$ fixed, with the aim of comparing  the performance of our algorithm with the reversible jump sampler of \cite{richgreen97} as implemented in  the \texttt{mixAK} R-package \citep{pacchettor}. Implementation of our algorithm has been done in C++ using the \texttt{Rcpp} library \citep{rcpp2011}, while post processing of the MCMC results in R. For each MCMC run, we have discarded the first 5000 iterations as burn-in and thinned every 10, obtaining a final sample size of $5000$. We have considered different  scenarios, and in Figure~\ref{fig:dens} we show the predictive density with 95\% credible bounds for one of them.

\begin{figure}[h!]
  \centering
  \includegraphics[scale=0.5]{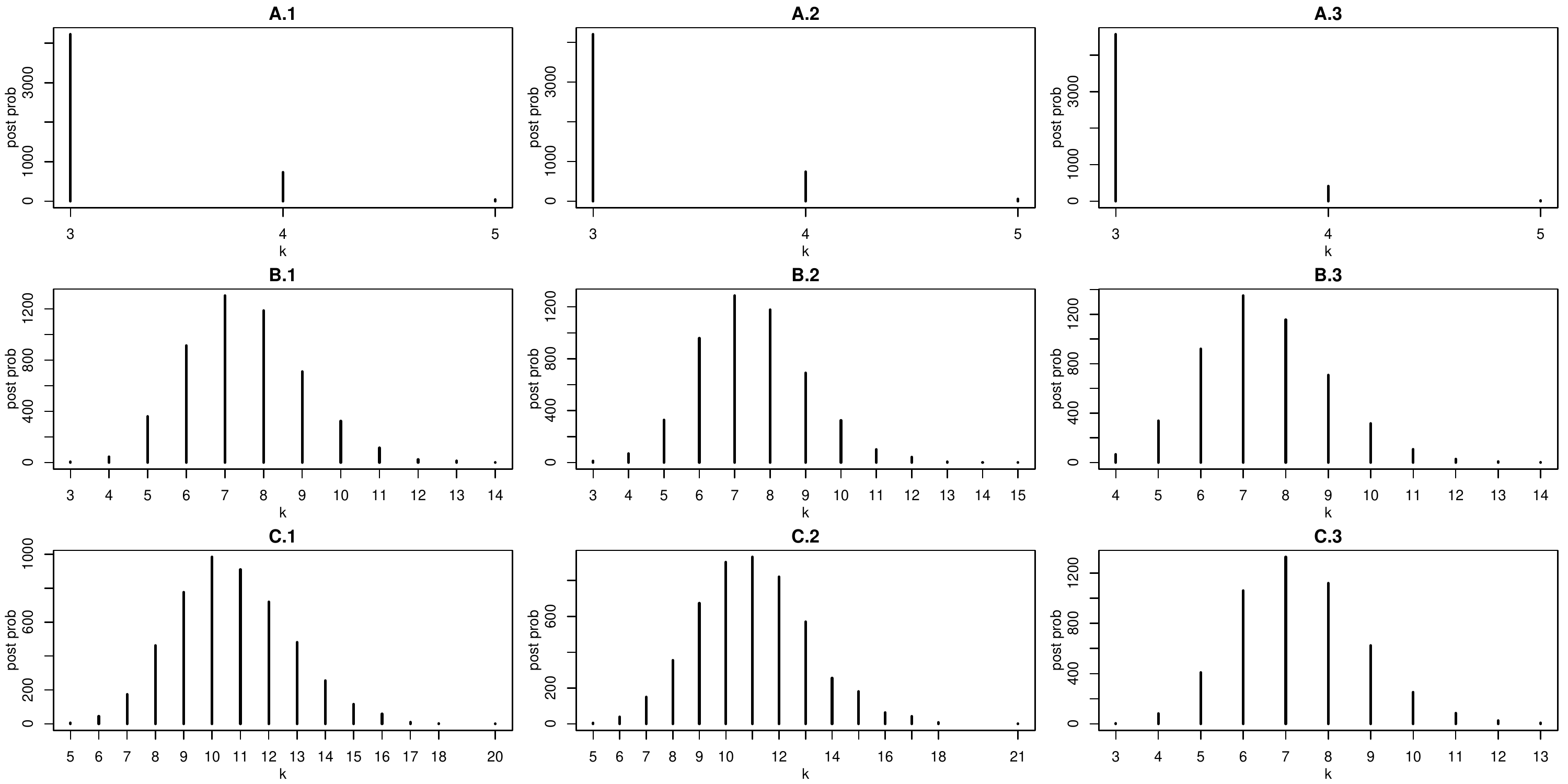}
  \caption{Posterior distribution of $k$ for the three scenarios.}
  \label{fig:post_k}
\end{figure}

\begin{figure}[h!]
  \centering
  \includegraphics[scale=0.5]{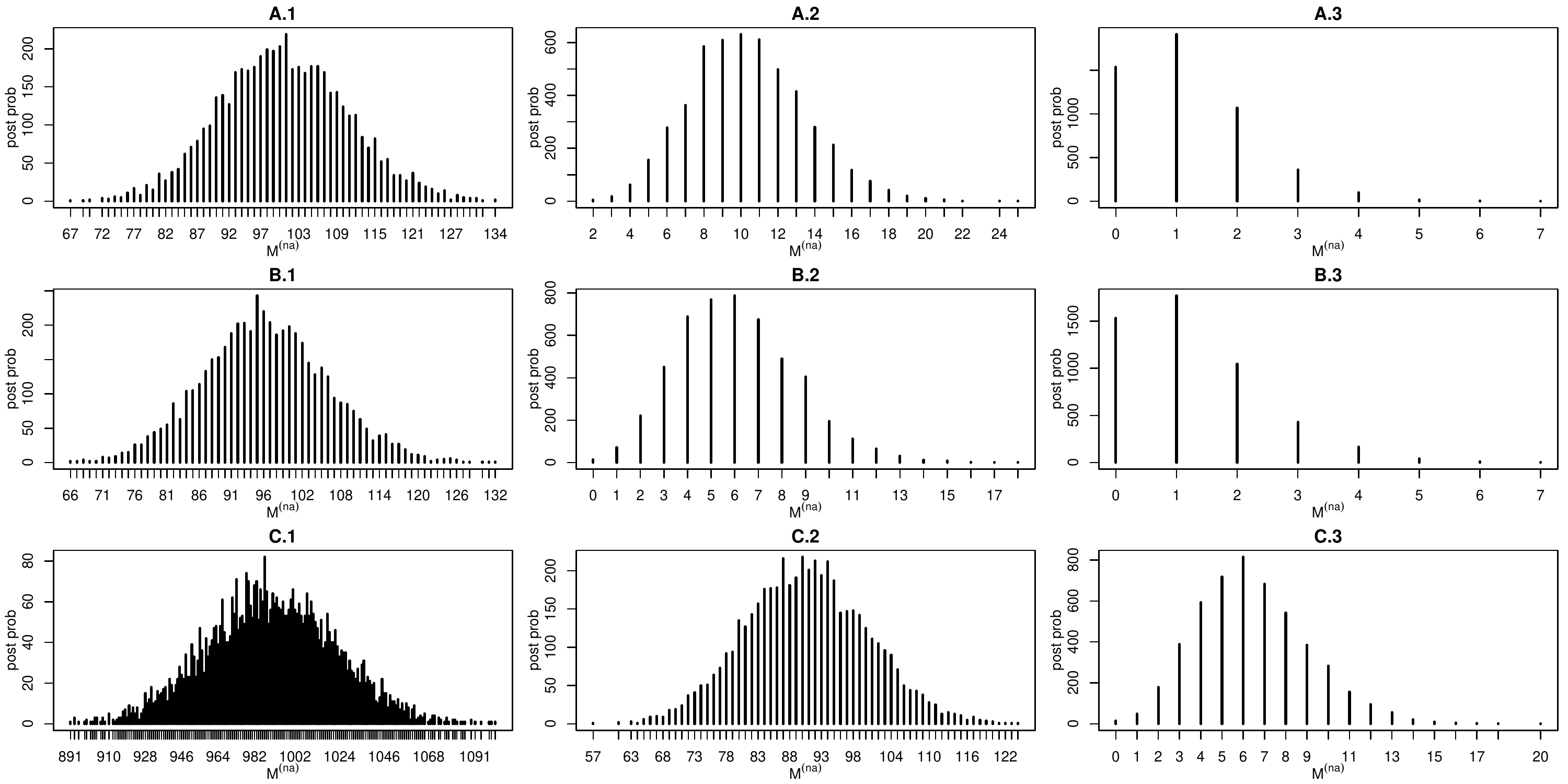}
  \caption{Posterior distribution  of $M^{(na)}$ for the three scenarios.}
  \label{fig:post_Mna}
\end{figure}

First of all, we  fix the hyperparameters $\gamma$ and $\Lambda$ in Eq.~\eqref{eq:latvar} in such a way that the prior mean for the number of clusters is (A) $\mathbb{E}(k)= 1$; (B) $\mathbb{E}(k)= 5$; (C) $\E(k)=10$.
In order to compare the conditional algorithm with the Reversible Jump, we compute the integrated autocorrelation time (IAC) and the effective sample
size (ESS) for the  number $M$ of components  for the all combinations of hyper-parameters. The IAC \citep{sokal1997monte} index provides a measure of the efficiency of the sampling algorithm in terms of accuracy of the estimates \citep[see, e.g.,][]{kalli2011slice}. A small absolute value of  the integrated autocorrelation time  (near 0) implies good  mixing and hence an efficient method. The Effective Sample Size \citep[ESS,][]{kong1992note} provides an estimate of the number of  independent draws from the posterior distribution  of a parameter of interest and small values indicate high autocorrelation between  draws, implying that  the estimate of the posterior distribution of that parameter will be poor. 
Posterior results are summarised in Table~\ref{tab:casi_A_B_C}: it is evident that  our algorithm outperforms the reversible jump in terms of both the IAC and ESS. 
\begin{table}
\caption{\label{tab:casi_A_B_C} Running times, posterior mean of $M$ and  integrated
autocorrelation times $\hat \rho$ for the Gibbs sampler (GS) in Appendix~\ref{sec:alg_fmm}
     and the Reversible Jump (RJ) MCMC implemented in the R-package  {\tt mixAK}.}
\centering
\fbox{%
\begin{tabular}{c|c||c|c|c||c|c|c}
  \  & \ & \multicolumn{3}{c||}{\textbf{GS}} & \multicolumn{3}{c}{\textbf{RJ}} \\
  \  & $(\Lambda,\gamma)$ & $\mathbb E (M|data)$ & ESS & IAC  &  $\mathbb E(M|data)$ & ESS $M$ & IAC $M$ \\
  \hline \hline
     \ &\ &\ &\ &\ &\ &\ &\ \\
                   &  $(100,2e^{-4})$   & 103.78  & 5474.59 & 1.49  & 70.48  &    3.07 & 602.26 \\
      \textbf{ A } &  $(10,2e^{-3})$    & 13.51   & 5000.00 & 1.51  & 11.43  &   16.38 & 149.08 \\
                   &  $(1,10e^{-2})$    & 4.22    & 1854.67 & 1.03  &  3.71  &  298.23 &   7.78 \\  
		   & & & & & & & \\
		   & $(100,1e^{-2})$    & 103.67  & 4231.39 & 0.58  & 95.84  &    8.66 & 206.10 \\
      \textbf{ B } & $(10,0.143)$       & 13.51   & 1551.56 & 1.38  & 10.18  &  467.97 &   5.77 \\
                   & $(5,0.5) $         & 8.69    & 1178.10 & 2.14  &  7.09  & 1168.48 &   1.99 \\ 
		   & & & & & & & \\
                   & $(1000,2.8e^{-3})$ & 1001.01 & 5393.03 & 1.50 & 819.49  &    1.94 & 874.78 \\
      \textbf{ C } & $(100,3.2e^{-2}) $ & 101.62  & 3846.80 & 0.57 & 86.23   &   22.55 &  86.35 \\
                   & $(10,1.8) $        & 13.71   & 1595.40 & 1.64 & 8.50    & 1271.37 &   2.03 \\ 
\end{tabular}}
\end{table}
Moreover, through an appropriate choice of $(\Lambda,\gamma)$, we are able to introduce in the model a desired level of sparsity. In Figure~\ref{fig:post_k} we report the posterior distribution of the number of clusters (allocated components) for the same combinations of hyper-parameters in Table~\ref{tab:casi_A_B_C}.
It is clear that the posterior distribution of the number of clusters is  robust to the choice of hyper-parameters  within each scenario (A, B and C),  since the prior mean on the number of allocated components is constant.
To gain more insight, in Figure~\ref{fig:post_Mna} we show the posterior distribution of $M^{(na)}$, the number of non-allocated components. We highlight: (i) these posteriors are more concentrate on large number for large values of $\Lambda$  (ii) for the same value of $\Lambda$ the level of sparsity increases for small values of $\gamma$ (see variations within columns). Large values of $\Lambda$ and small values for $\gamma$ favour  a posterior distribution for $M^{(na)}$ centred on large values.
We conclude that $\Lambda$ controls the number of unallocated clusters, while $\gamma$ controls degree of  sparsity of the mixture. 

\begin{figure}[h!]
  \centering
  \includegraphics[scale=0.27]{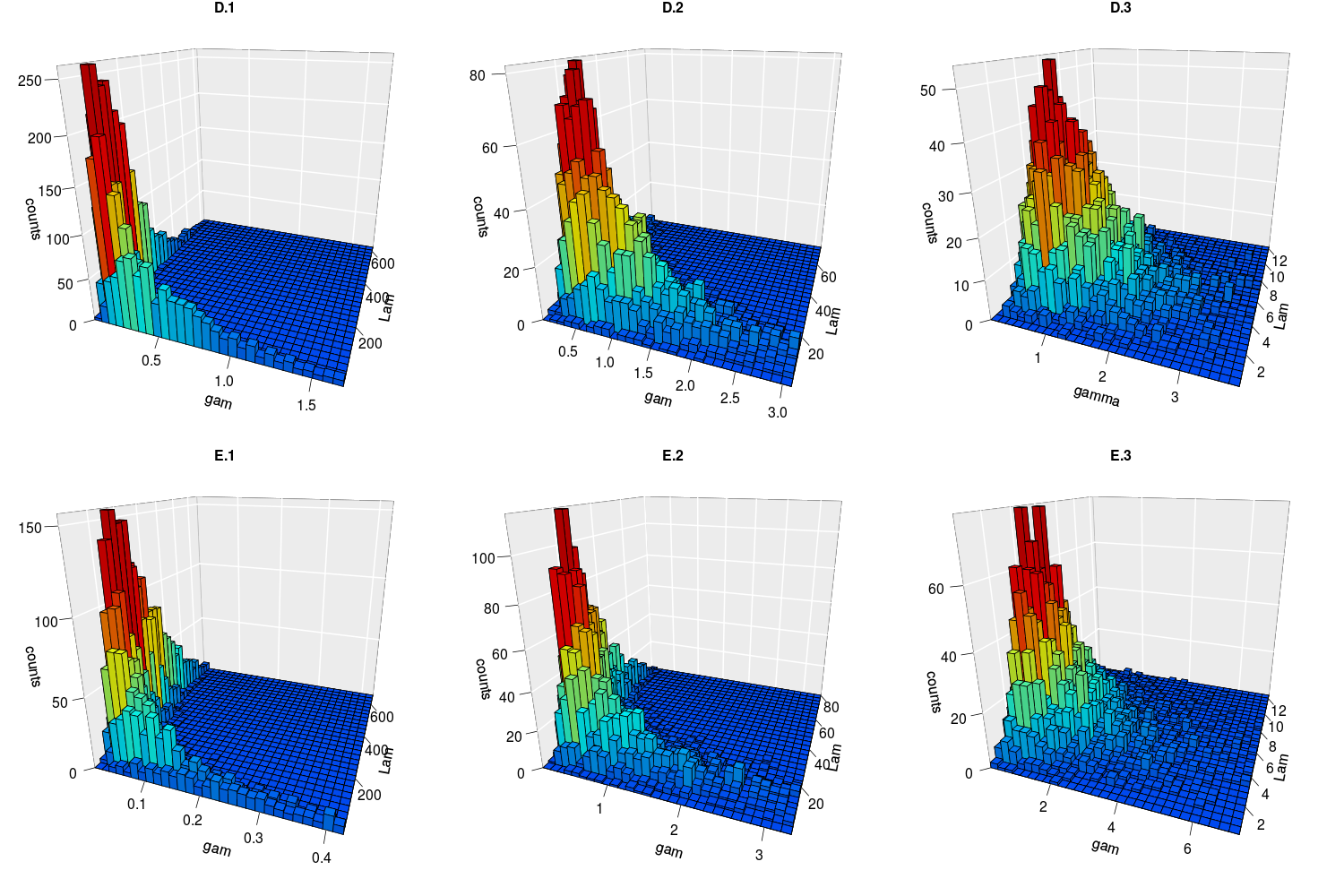}
  \caption{Three-dimensional histogram of the joint marginal posterior of $\Lambda$ and $\gamma$ for scenarios D and E.}
  \label{fig:3dhis_lam_gam}
\end{figure}

Finally, we fit the same model adding an extra layer to the  hierarchy, by specifying  prior distributions on  both $\gamma$ and $\Lambda$. We consider two scenarios  described in Table~\ref{tab:casi_D_E}. In this case we consider the posterior distribution of the number of allocated components, that is more meaningful from an inferential point of view, as well as a predictive goodness of fit criterion, the Logarithm of the  Pseudo Marginal Likelihood \citep[LPML --][]{geisser1979predictive}.
\begin{table}
\caption{\label{tab:casi_D_E} Results obtained specifying a prior  on $\Lambda$ and $\gamma$.}
\fbox{%
\begin{tabular}{c|c||c|c|c|c|c|c|c}
     \            & $\Lambda$            &   $\gamma$ &  $\mathbb E (k|data)$ & ESS $k$ & IAC $k$  &  LPML \\
      \hline \hline
      & & & & & &  \\
                   &  Gamma$(1,0.01)$   & Gamma$(1,1)$  &  13.41 &  909.07  &  3.39 &  -1632.44  \\
      \textbf{ D } &  Gamma$(1,0.1)$    & Gamma$(1,1)$  &  11.73 &  723.87  &  3.65 &  -1515.55  \\
                   &  Gamma$(1,1)$      & Gamma$(1,1)$  &   7.72 &  307.77  &  10.38 &  -1092.03  \\  
		   & & & & & &  \\
		   & Gamma$(1,0.01)$    & Gamma$(0.1,0.1)$  & 11.79 &  454.57  &  5.20 &  -1471.48  \\
      \textbf{ E } & Gamma$(1,0.1)$     & Gamma$(0.1,0.1)$  & 11.41 &  397.18  &  7.15 &  -1443.14  \\
                   & Gamma$(1,1)$       & Gamma$(0.1,0.1)$  &  7.89 &  199.44  &  12.79 &  -1139.50  \\ 
    \end{tabular}}
\end{table}
In Figure~\ref{fig:post_k_random} of Appendix~\ref{sec:add_figures} we show the posterior distribution of number of clusters: we note how adding an extra layer to the hierarchy makes inference more robust to hyper-prior specifications.
In Figure~\ref{fig:post_Mna_random} of Appendix~\ref{sec:add_figures} the posterior distribution of non-allocated components is shown: it is evident that $\Lambda$ still  influences such distribution, while $\gamma$ determines the level of sparsity. Moreover, adding this extra level of randomness induces more parsimonious posteriors: the posterior on the number of non-allocated components is now shrunk toward  zero in both scenarios D and E. Although treating  $\Lambda$ and $\gamma$ as random variables leads to more robust estimates, it also increases the autocorrelation in the MCMC chains. This is evident from Table~\ref{tab:casi_D_E} as well as from Figure~\ref{fig:3dhis_lam_gam} that shows the joint marginal posterior of $\Lambda$ and $\gamma$ for scenarios D and E.  We highlight the strong negative correlation between the two hyper-parameters, which is natural as $\Lambda$ controls the number of on non-allocated components while $\gamma$ the number of allocated ones.  

\section{Population Structure: Taita Thrush Data}
\label{sec:popstruct}
In population genetics, population structure refers to the presence of a systematic differences in genetic markers' allele frequencies between subpopulations due to variation in ancestry. This phenomenon arises from the bio-geographical distribution of species, due to the fact that either natural populations occupy a vast geographic area and cannot act as randomly mating or geographical barriers reduce migration between different regions. Consequently population structure affects  the dynamics of alleles in populations and impacts the type of statical analysis to perform in many  applications, for example in genetic association studies.   Broadly speaking, the analysis of population structure focuses on: (i) detecting population structure in a sample of chromosomes; (ii) estimating the number of populations in a sample; (iii) assigning individuals to populations and (iv) defining the number of ancestral populations in a sample.
A variety of statistical approaches have been proposed to infer population structure. Arguably the most widely method  is the one proposed by \cite{pritchard2000inference} based on Bayesian mixture models and  implemented in the software {\tt STRUCTURE} \citep{pritchard2003documentation}. 
 \cite{pritchard2000inference} assume that individuals come from one of $M$ (fixed)  subpopulations and population membership and population specific allele frequencies are jointly estimated from the data. Independent priors on the allelic profile parameters of each population are specified and posterior inference is performed through MCMC.  
In \cite{pritchard2000inference}, the number of mixture components is fixed and their method clusters individual in one of a fixed number of populations.  Determination of the number of populations in a sample is  achieved using a model selection criteria based on MCMC estimates of the log marginal probabilities of the data and the Bayesian deviance information criterion, though it has been noted by \cite{falush2003inference} that such estimates are highly sensitive to prior specifications regarding the relatedness of the populations. 
To avoid such model selection, \cite{huelsenbeck2007inference} propose a method for the analysis of population structure based on a Dirchlet process mixture model and implemented in the software {\tt Structurama} \citep{huelsenbeck2011structurama}, which  does not require the specification of a fixed and finite $M$.

\begin{figure}[h!]
\centering
\includegraphics[width=0.7\textwidth]{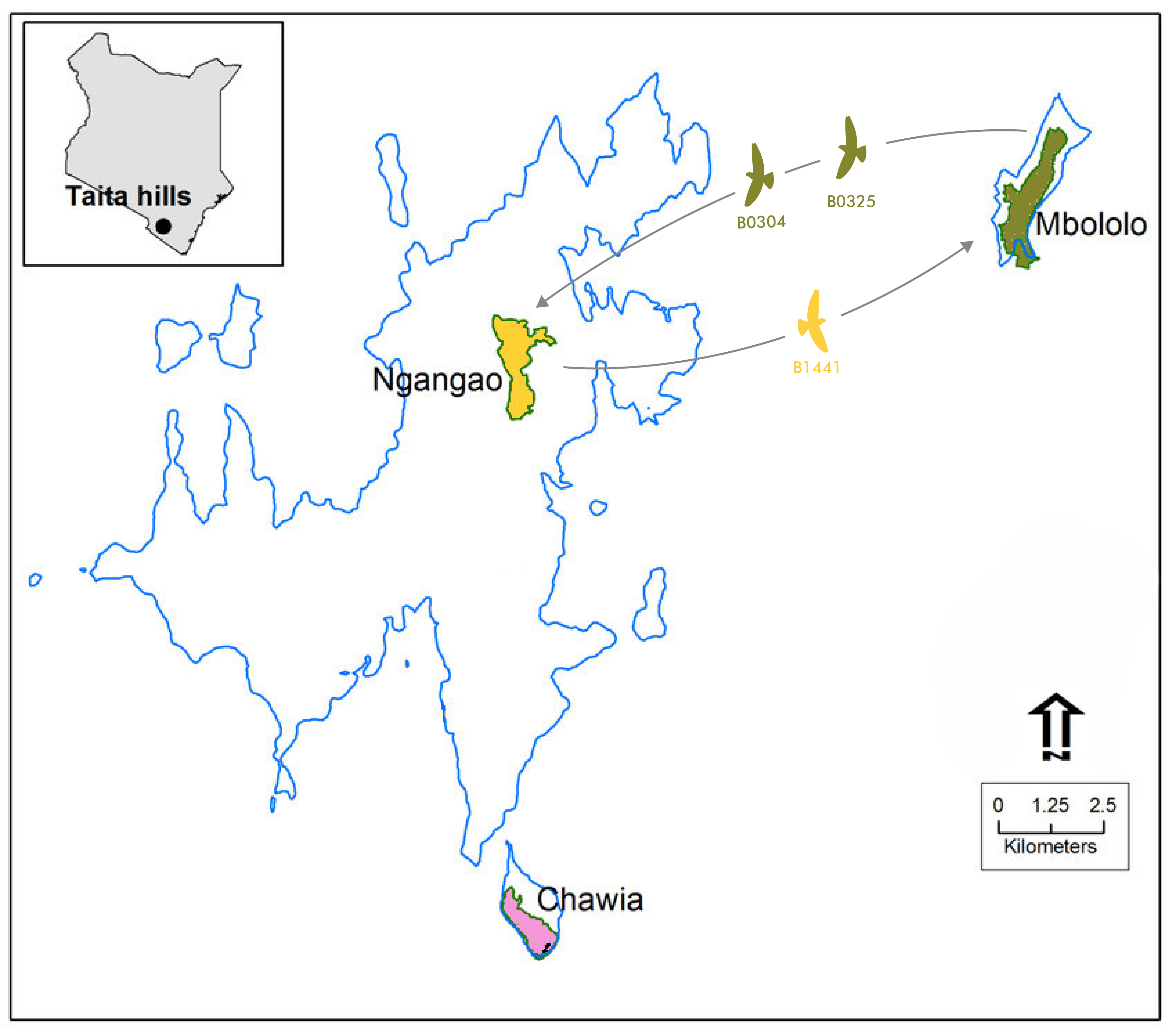}
\caption{Posterior estimate of the clustering allocation: each colour correspond to a cluster. Note that he two green thrushes have been captured in Ngangao, but have the genetic profile of the Mbololo birds. The opposite is true for thrush B1441.}
\label{fig:tordimigranti}
\end{figure}

We now illustrate the performance of our method in a population structure problem, using an empirical data set of $n=237$ Taita thrushes kindly made available by Dr P. Galbusera. A previous smaller version of these data  \citep{galbusera2000genetic}  has been analysed by \cite{pritchard2000inference} and \cite{huelsenbeck2007inference} as benchmark example. We have run an analysis using our algorithm on this old data and have drawn identical conclusions. Here we prefer to focus on the new dataset.  
The Taita Hills in Kenya represent the northernmost part of the Eastern Arc Mountains biodiversity hotspot of Kenya and Tanzania. They are isolated from other highlands by over 80 km of semiarid plains in either direction. During the last 200 years, indigenous forest cover in the Taita Hills has decreased by circa 98\% and
the critically endangered  Taita thrush, endemic to the Taita Hills, is currently restricted to the fragments of Mbololo, Ngangao 
and Chawia \citep{callens2011genetic}.
These fragments are separated from each other by cultivated areas and human settlements.  This dataset is ideal to test the performance of our method as the geographic samples are likely to represent distinct populations, i.e. mixture components.       
Each   bird was sampled at $L=6$ microsatellite loci. We follow the notation of \cite{huelsenbeck2007inference}. Recall that the Taita thrush is diploid, i.e. has two sets of chromosomes and for each locus we have genotype data.   
At locus $l$, we observe $J_l$ unique alleles. The number of
copies of allele $j$ at locus $l$ in individual $i$ is denoted by
$Y_{ilj}\in\{0,1,2\}$
   and  the number of copies of all alleles observed at locus
$l$ in individual $i$ is denoted by $Y_{il}=\sum_{j=1}^{J_l}Y_{ilj}$.    
    The allelic information for individual $i$ at locus $l$ is
contained in the vector
${Y}_{il}=(Y_{il1},Y_{il2},\dots,Y_{ilJ_{l}})$, with the constrain
$\sum_{j=1}^{J_l}Y_{ilj}=2$. 
Given $M$ possible populations, let $\tau_{mlj} $ denote the frequency of allele $j$ al locus $l$ in population $m$, let $\tau_{ml} =(\tau_{ml1},\ldots, \tau_{mJ_l} ) $ be the vector of allele frequencies at locus $l$ in population $m$ and let $\tau_m = (\tau_{m1},\ldots, \tau_{mL}) $. Finally, let $c_i \in \{1, \ldots,M\}$ be the allocation variable of bird $i$, i.e. $c_i=m$  if the bird comes from population $m$. Following \cite{huelsenbeck2007inference} we assume that
$$
f_{}\left( {y}_{il}|{\tau}_{ml} \right)=\Prob({Y}_{il}={y}_{il}\mid {\tau}_{ml}, c_i=m)\propto \prod_{j=1}^{J_l}\tau_{mlj}^{y_{ilj}}$$
We assume independence across loci, so that, if
${Y}_i=({Y}_{i1},\dots,{Y}_{iL})$ is the multidimensional array of the
allelic information at the $L$ loci for individual $i$, we have 
    \begin{equation}
     f_{}\left( {y}_{i}\mid {\tau}_{m},c_i=m
\right)=\Prob({Y}_{i}={y}_{i} \mid {\tau}_{m},c_i=m)=\prod_{l=1}^{L}
    f_{}\left( {y}_{il}\mid {\tau}_{ml} \right)
      \label{eq:sampling_pop_struc}
    \end{equation}
    We fit Model \eqref{eq:latvar}, with the sampling model defined in Eq.~\eqref{eq:sampling_pop_struc}. The  mixing measure is  a finite Dirchlet process as in Section \ref{sec:FDP}, with the following prior specification: $M$ has a shifted Poisson prior distribution with parameter $\Lambda$, $P_0$ is the convolution of $L$ independent Dirichlet distributions with parameter 1, $\gamma$ in the finite Dirichlet process has a Gamma prior with parameter $(0.1,0.1)$, $\Lambda$ has a Gamma prior with parameter $(3/2,1/2)$. For the parameter $\gamma$ we have specified a vague prior distribution, while  the hyper-parameters in the prior for $\Lambda$ are chosen so that the prior  mean is 3, corresponding to the three geographical fragments, and the prior variance is large. We employ the conditional algorithm described in  Appendix~\ref{sec:alg_fmm} to perform posterior inference. The mode of the posterior distribution for $k$ is at 3 ($\E(k\mid data) = 3, \Var(k\mid data)=0.03$), as well as the one of the posterior of $M$ ($\E(M\mid data) = 3.12, \Var(M\mid data)=0.42$). 
    From Figure~\ref{fig:tordimigranti} it is evident that the three  clusters coincide with the  three geographical fragments, except in three cases where the birds appear to be out of the obvious clusters. This could be due to rare migration events \citep{galbusera2000genetic}.

\begin{figure}[h!]
  \centering
   \includegraphics[width=\textwidth]{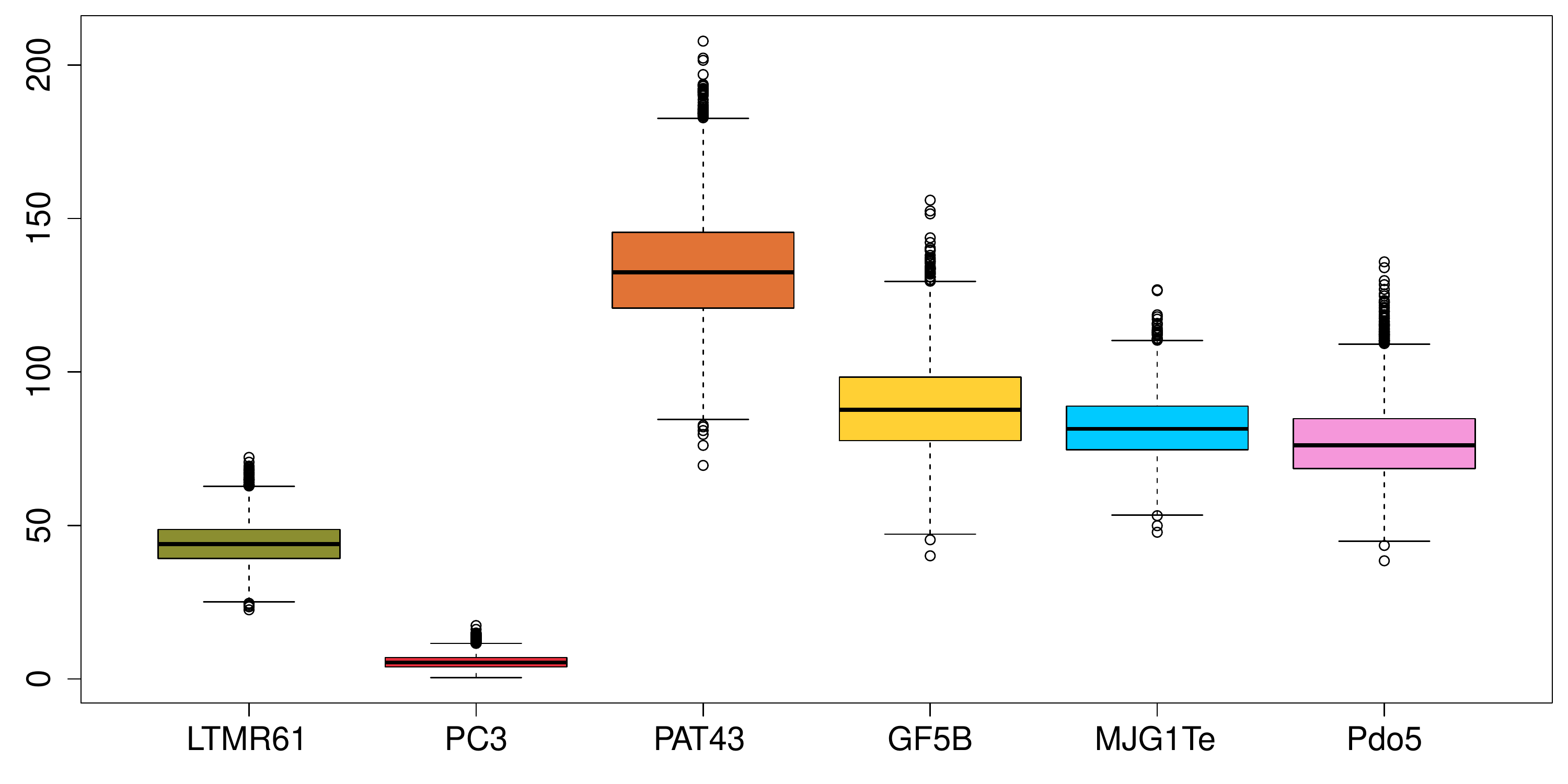}
\caption{Posterior distribution of the KL divergence for each microsatellite locus.}
   \label{fig:KL}   
\end{figure}

An important goal of population structure  analysis is not only to uncover the group structure of the observations, but also to identify variables that best distinguish the different populations. The results could lead to a better understanding of the evolutionary patterns of population differentiation.  To this end we would like to identify the microsatellite loci that most influence the clustering structure. Variable selection for clustering is a challenging problem since there is no observed response to inform the selection and  the inclusion of unnecessary variables could complicate or mask the recovery of the clusters \citep{tadesse2005bayesian,kim2006variable}. As such there are few contributions in the literature. Here we opt for a  model choice method proposed by \cite{goutis1998model} in the generalised linear model framework, which we adapt to our context. The approach of \cite{goutis1998model} focuses on the predictive properties of a model and, employing the Kullback-Leibler distance as discrepancy measure,  aims to assess the relevance of some restriction on the parameter $\Theta$ (leading to a simpler model) with respect to a full model described by a density $f(y \mid \theta)$. More in details, for each locus $l$, let $Y_l=(Y_{1l},\ldots, Y_{nl})$, $\theta_{il}=\tau_{ml}$ if $c_i=m$ and $\theta_l =(\theta_{1l},\ldots, \theta_{nl})$. Let $f(y_l\mid \theta_l)$ be the full general mixture model:
\begin{equation}
\label{eq:fullmdel}
f(y_l\mid \theta_l)\propto \prod_{i=1}^{n} 
    \prod_{j=1}^{J_l} \theta_{ilj}^{y_{ijl}}
\end{equation}     
We define  a model choice hypothesis $H_0$ through a restriction on the parameter space, i.e. $\theta_l\in \Theta_0\subset \Theta $, where $\Theta_0$ is the subset of the parameter space such that $\theta_{ilj} =\widetilde{\theta}_{lj}$ for each $i$. In our application $H_0$ represents a fuly parametric model for locus $l$. \cite{goutis1998model} define the projection $\theta_l^\perp$ of $\theta_l$ according to the Kullback-Leibler distance $d$ to be the point in $\Theta_0$ that achieves the infimum
$$
 d \left\lbrace f(\cdot \mid \theta_l), f(\cdot \mid \theta_l^\perp) \right\rbrace = \inf_{\widetilde{\theta}_l \in \Theta_0} d \left\lbrace f(\cdot \mid \theta_l), f(\cdot \mid \widetilde{\theta}_l) \right\rbrace 
$$ 
where $\widetilde{\theta}_l =(\widetilde{\theta}_{l1}, \ldots, \widetilde{\theta}_{lJ_l})$ and $   f(\cdot \mid \theta_l^\perp)$ is the projection of  $f(\cdot \mid \theta_l)$. Obviously  small values of $d$ support $H_0$.
We opt for this approach because, instead of phrasing the problem in terms of the classical dichotomy between null and alternative hypothesis, it interprets model choice in terms of the approximation efficacy of a more parsimonious model, focusing on whether or not $\theta_l$ is far away from the subspace $\Theta_0$.  In Figure~\ref{fig:KL}
 we show the posterior distribution of $ d \left\lbrace f(\cdot \mid \theta_l), f(\cdot \mid \theta_l^\perp)\right\rbrace $ for each locus $l$. It is evident that locus PC3 contributes the least to the clustering structure as the distance is concentrated near zero, implying the its allele frequencies are similar across Taita thrush populations. The other loci, in particular PAT43, present  allele frequency differences among the three groups, which in our case well correspond  to geographical locations.

  \section{Conclusions}
  \label{Conclusions} 
 In this work we contribute to the growing understanding of mixture models by providing an unifying framework which encompasses both finite and infinite mixtures.
 A key concept is the 
 distinction between the number of components and number of clusters, where by components we refer to the number of subpopulations that are likely to have generated the data, while clusters indicate the number of non-empty components in a sample. Already \cite{nobile2004posterior} had pointed out this difference, noticing that the posterior distribution of the number of components $M$ (corresponding to the data generating process) might assign considerable probability to values greater than $k$, the number of clusters.  
More recently, the concept of \textit{sparse finite mixtures} has been introduced as a first attempt to bridge between finite mixture models and nonparametric mixtures   
\citep[see][]{malsiner2016model,malsiner2017identifying}. 
 In this context, \cite{fruhwirth2017here} fix $M$ very large so that they are \textit{close to infinity}, justifying this choice from the asymptotic point of view and then their work focuses on sparse estimation of the number of clusters.  Our construction is based on the normalization of a point process, which is a standard trick in Bayesian nonparametrics. We  introduce the Norm-IFFP  prior process and we provide theoretical results characterizing the induced prior on the partition of the observations and  the posterior distribution of this process. 
  Our framework allows for efficient computations (inherited  from the nonparametric construction) and for data driven estimation of both number of clusters and components,  as well as of any functional of interest.

 \section*{Acknowledgement}
 We would like to thank Dr Peter Galbusera at the Royal Zoological Society of Antwerp for sharing the enriched  Taita Thrush Dataset. Dr Argiento is grateful to Yale-NUS College, Singapore for the funding provided.

\appendix
\renewcommand{\theequation}{\thesection.\arabic{equation}}
\renewcommand{\thefigure}{\thesection.\arabic{figure}}

\section{Appendix: Proofs}

\subsection{Proof of Theorem \ref{thm:eppf_peps}}
\label{sec:app_proof_eppf}

\begin{proof}
 We have
\begin{eqnarray}
\pi(n_1,\dots,n_k)=\sum_{m=1}^{+\infty} \pi(n_1,\dots,n_k|M=m)q_m
\label{eq:eppf_sum} 
\end{eqnarray}
since $M\sim q_M$ and we have assumed $q_0=0$. Then, equation (3) in \cite{pitman2003poisson} yields
\begin{equation*}
\pi(n_1,\dots,n_k|M=m)=  \uno_{ \{k\le m\}}
\sum_{(c_1^\star,\dots,c_k^{\star})}^{}\E\left(
  \prod_{j=1}^{k}w_{c_j^\star}^{n_j} \right)
\end{equation*} 
where the vector $(c_1^{\star},\dots,c_k^{\star})$ ranges over all permutations of $k$ positive integers in $\{1,\dots,m\}$.  Recall that  $w_m=\frac{S_m}{T}$ as defined in Eq.\eqref{eq:nor_ifpp}. Under the assumptions  in Section~\ref{sec:ffp} and Section~\ref{sec:nifpp} 
the joint law of the unnormalized jumps of $P$ is
$\mathcal{L}(dS_0,\dots,dS_{m}|M=m)=\prod_{l=1}^{m}h(S_l)dS_l$. Then, using the identity
 $1/T^n=\int_{0}^{+\infty}1/\Gamma(n)u^{n-1} e^{-u T} du$,
we have: 
 \begin{align*}
    \pi(n_1,..,n_k|M=m) &=\uno_{\{k\le m\}} \sum_{(c_1^\star,\dots,c_k^\star)}    \int_{}^{}
   \prod_{j=1}^{k}\frac{S_{c_j^\star}^{n_j}}{T^{n_j}} \mathcal{L}(dS_1,\dots,dS_{m}|m)\\
   &= 
\uno_{\{k\le m\}} \sum_{(c_1^\star,\dots,c_k^\star)} \int_{0}^{+\infty} \left( \frac{1}{\Gamma(n)}u^{n-1}  
\prod_{j=1}^{k}\int_{0}^{+\infty}S_{c_j^\star}^{n_j}\e^{-S_{c_j^\star}u}h(S_{c_j^{\star}})dS_{c_j^{\star}} \right.   \\ 
  &  \qquad \qquad \qquad \qquad  \qquad \qquad \qquad \qquad
  \times \left.
  \prod_{c\notin(c_1^*,\dots,c_k^\star)}^{}\int_{0}^{+\infty}\e^{-S_{c}u}h(S_c)dS_c \right) du\\
&= 
\uno_{\{k\le m\}} \sum_{(c_1^\star,\dots,c_k^\star)} 
\int_{0}^{+\infty}  \frac{1}{\Gamma(n)}u^{n-1}  
\prod_{j=1}^{k}
\E\left( S_{c_i^\star}^{n_j}\e^{-S_{c_j^\star}u} \right)
\prod_{c\notin(c_1^*,\dots,c_k^\star)}^{}\E(\e^{-S_c u}) du
\\ 
  &= 
\uno_{\{k\le m\}}
 \sum_{(c_1^\star,\dots,c_k^\star)} 
\int_{0}^{+\infty}  \frac{1}{\Gamma(n)}u^{n-1}  
\prod_{j=1}^{k}\kappa(n_j,u)
 \psi(u)^{m-k}du
\\ 
 &= 
\uno_{\{k\le m\}}
\frac{m!}{(m-k)!}
\int_{0}^{+\infty}  \frac{1}{\Gamma(n)}u^{n-1}  
\prod_{j=1}^{k}\kappa(n_j,u)
 \psi(u)^{m-k}du 
 \end{align*}
 where  $\psi(u)=\int_{0}^{\infty}e^{-us}h(s)ds$ is the Laplace transform of the density $h(s)$, while for each integer $n$,
    $\kappa(n_,u):=\int_{0}^{\infty}s^{n}\e^{-u s}h(s)ds=
    (-1)^{n}\frac{d}{du^{n}}\psi(u)
    $. We exploit the fact that in the second last equation, the summation term does not depend on the indices $(c_1^\star,\dots,c_k^\star)$, with   $\sum_{c_1^\star,\dots,c_k^\star}1=\frac{m!}{(m-k)!}$ since   $(c_1^\star,\dots,c_k^\star)$ ranges over all permutations of $k$ positive integers  between $1$ and $m$.

\noindent Now, combining Eq.~\eqref{eq:eppf_sum} with the equality derived above, we obtain
\begin{align*}
 \pi(n_1,\dots,n_k)&=\sum_{m=1}^{+\infty} \pi(n_1,\dots,n_k|M=m)q_m\\
&=\sum_{m=1}^{+\infty}
\uno_{\{k\le m\}}
\frac{m!}{(m-k)!}
\int_{0}^{+\infty}  \frac{1}{\Gamma(n)}u^{n-1}  
\prod_{j=1}^{k}\kappa(n_j,u)
\psi(u)^{m-k}du\\
&=\sum_{m=k}^{+\infty}
\frac{m!}{(m-k)!}
\int_{0}^{+\infty}  \frac{1}{\Gamma(n)}u^{n-1}  
\prod_{j=1}^{k}\kappa(n_j,u)
\psi(u)^{m-k}du\\
&=\sum_{m=0}^{+\infty}
\frac{(m-k)!}{(m)!}
\int_{0}^{+\infty}  \frac{1}{\Gamma(n)}u^{n-1}  
\prod_{j=1}^{k}\kappa(n_j,u)
\psi(u)^{m}du
\end{align*}
which gives Eq.~\eqref{eq:eppf_expr} and concludes the proof.

  \end{proof}

\subsection{Proof of Corollary \ref{cor:priorK}}
\label{sec:app_proof_priorK}
\begin{proof}

Let $(n_1,\dots,n_k)$ be a composition of $n$, i.e.  a vector of $k$ positive integers $(n_j>0\ i=1,\dots,k)$ such that
$n=\sum_{j=1}^{k}n_i$. There are 
$${ {n}\choose{n_1,\dots,n_k}}\frac{1}{k!}$$
partitions of $\{1,\dots,n\}$ such that their cluster size is given by  with $(n_1,\dots,n_k)$.
Thus  it is clear that  the joint prior probability of sampling a partition (clustering configuration) with $k$ clusters and 
cluster sizes $(n_1,\dots,n_k)$ is 
\begin{equation*}
  { {n}\choose{n_1,\dots,n_k}}\frac{1}{k!}\pi(n_1,\dots,n_k)
\end{equation*}
The marginal prior probability of sampling a partition with $k$ cluster is computed  summing over 
all the possible cluster sizes $(n_1,\dots,n_k)$:
 \begin{align*}
   p^\star_k&=\sum_{n_1,\dots,n_k}^{} { {n}\choose{n_1,\dots,n_k}}\frac{1}{k!}\pi(n_1,\dots,n_k)\\
   &=  
    \int_{0}^{+\infty}
    \dfrac{u^{n-1}}{\Gamma(n)}
    \left\{ 
      \sum_{m=0}^{\infty}\frac{(m+k)!}{m!}\psi(u)^{m}q_{m+k}
    \right\}
\frac{n!}{k!} \sum_{n_1,\dots,n_k}
    \prod_{j=1}^k 
\frac{
\kappa(n_j,u)du}{n_j!}  \\
&=  
    \int_{0}^{+\infty}
    \dfrac{u^{n-1}}{\Gamma(n)}
    \left\{ 
      \sum_{m=0}^{\infty}\frac{(m+k)!}{m!}\psi(u)^{m}q_{m+k}
    \right\}
    B_{n,k}(\kappa(\cdot,u))
 \end{align*}
 where $B_{n,k}(\kappa(\cdot,u))$ is the \emph{partial Bell polynomial} \citep{pitman2006combinatorial} for the sequence of coefficient $\{\kappa(n,u),\ n=1,2,\dots\}$.

\end{proof}

\subsection{Proof of Theorem \ref{theo:Posterior}}
\label{sec:app_proof_post_char}

\begin{proof} 
  Let  $  \boldsymbol \theta=(\theta_1,\dots,\theta_n)$
  be a sample from $P\sim \text{Norm-IPPF}(h,p_0,q_M)$:
  \begin{eqnarray*}
  \theta_1,\dots,\theta_n \mid  P&\iid & P\\
  P &\sim &\text{Norm-IPPF}(h,p_0,q_M)
\end{eqnarray*}

With a slight abuse of notation, we use $\mathcal{L}(X)$ to denote the pdf or pmf of a random variable $X$ and $\mathcal{L}(\tilde P)$ to indicate the Janossy density of a point process. We need to show that the posterior density 
  $\calL(\tilde P|\theta_1,\dots,\theta_n)$ is still the Janossy density of a finite
  point process. Indeed,
  \begin{equation}\label{eq:post_1}
    \begin{split}
      \mathcal{L}(\tilde P|\boldsymbol{\theta} )&\propto
      \mathcal{L}(\boldsymbol{\theta}, \tilde P)
      =\calL(\theta_1,\dots,\theta_n \mid\tilde P)\calL(\tilde P)
      =\prod_{i=1}^{n}P(\theta_i)\calL(\tilde P)\\
      &= \left\{  \frac{1}{T^n}\prod_{i=1}^{n}\left( \sum_{m \in \cal M}
      S_m\delta_{\tau_m}(\theta_i)
    \right)  \right\}
    M! p_M \prod_{m\in \cal M}h(S_m)p_0(\tau_m)\\
  \end{split}
\end{equation}
We introduce the variable $U_n\in \Rea^+$ such that the 
$\mathcal{L}(U_n \mid T)$ is a Gamma$(n,T)$ and whose marginal distribution is given by Eq.~\eqref{eq:margU}. Recall that
\begin{eqnarray*}
T&= &\sum_{m\in \cal M} S_m\\
 \frac{1}{T^n}	&= & \int_0^n \frac{1}{\Gamma(n)} u^{n-1}e^{-uT} \text{d} u  \\
\end{eqnarray*}
Then
\begin{equation}\label{eq:post_2}
  \begin{split}
    \mathcal{L}(\tilde P,u |\boldsymbol{\theta} )&\propto
    \frac{1}{\Gamma(n)}u^{n-1}\e^{-u T}
    \left\{ \prod_{i=1}^{n}\left( \sum_{m \in \cal M}
    S_m \delta_{\tau_m}(\theta_i)
  \right)  \right\}
  M! q_M \prod_{m \in \cal M}h(S_m)p_0(\tau_m)\\
  &=
  \frac{1}{\Gamma(n)}u^{n-1}\e^{-u 
    \sum_{m\in \cal M} S_m
  }
  \left\{ \prod_{i=1}^{n}\left( \sum_{m\in \cal M}
  S_m\delta_{\tau_m}(\theta_i)
\right)  \right\}
M! q_M \prod_{m\in \cal M}h(S_m)p_0(\tau_m)\\
&=
\frac{1}{\Gamma(n)}u^{n-1}
\left\{ \prod_{i=1}^{n}\left( \sum_{m \in \cal M}
S_m\delta_{\tau_m}(\theta_i)
\right)  \right\}
M! p_M \prod_{m \in \cal M}
\e^{-u S_m}
h(S_m)p_0(\tau_m)\\
\end{split}
\end{equation}
Since $(\theta_1,\dots,\theta_n)$ is a sample from a discrete distribution, there is a positive probability of ties among the $\theta_i$s.
We  denote with $\bm \theta^\star=(\theta^\star_1,\dots,\theta^\star_{k})$ the vector
of distinct values, with
$k\le \min\{M,n\}$. Moreover we denote with $(n_1,\dots,n_k)$ of induced clusters in the sample $(\theta_1,\dots, \theta_n)$, i.e.
$n_j=\#\{i:\theta_i=\theta^{\star}_j\}$, $j=1,\dots,k$.
\newline Let ${\cal M}^{(a)}$ be the set of  indexes of the allocated points, $k=M^{(a)} \leq M$. Moreover,  conditionally on $M$, the process $\widetilde{P}=\{ (S_1, \tau_1),\ldots, (S_M, \tau_M)\}$ is defined on an $M$-dimensional space. Note that $M=M^{(a)}+ M^{(na)}$. Without loss of generality and for ease of notation, we assume that the set ${\cal M}^{(a)}$ is ordered, i.e. ${\cal M}^{(a)}=\{1,\dots,k=M^{(a)}\}$. This assumption is  inconsequential as ${\cal M}^{(a)}$ is a set of indices and the names of the labels are irrelevant. If $m \in {\cal M}^{(a)}$, then $\tau_m = \theta^\star_m$ and there is a one to one correspondence between the set $\{ \theta_m^\star, m=1,\ldots,k\}$ and $\{ \tau_{m},m=1,\ldots,M^{(a)} \}$.     
\noindent Let 
${\cal M}^{(na)}$ be the set of indexes corresponding to unallocated jumps.
A posteriori, conditionally on $(\theta_1,\ldots, \theta_n)$, ${\cal M}^{(na)}=\{k+1,\ldots, M\}$ and   $\widetilde P=\widetilde
P^{(a)}\cup \widetilde P^{(na)}$. If $M=k$, then ${\cal M}^{(na)}$ coincides with the  empty set. 
Conditionally on $u$, we can now obtain the Janossy measure of the posterior distribution:
\begin{eqnarray}
    \mathcal{L}(\widetilde P|u,\boldsymbol{\theta} )&\propto &  \mathcal{L}(\widetilde P, u\mid \boldsymbol{\theta} ) \nonumber  
  \\ 
  &\propto & \left\{ \prod_{ m \in \calM^{(a)}}
    S_m^{n_m} 
    \delta_{\tau_m}(\theta^\star_m)  \right\}
    (M^{(na)}+k)! q_{M^{(na)}+k}  \prod_{m\in \calM}
    \e^{-u S_m}
    h(S_m)p_0(\tau_m)  \nonumber \\
   &= &
    \left\{ \prod_{m\in \calM^{(a)}}
    \e^{-u S_m}S_m^{n_m}
    \delta_{\tau_m}(\theta^{\star}_{m})
  h(S_m)p_0(\tau_m)\right\} 
   (M^{(na)}+k)! q_{M^{(na)}+k} \nonumber
    \\  & & \cdot \prod_{m\in \calM^{(na)}}
    \e^{-u S_m}
    h(S_m)p_0(\tau_m) \nonumber \\
    &= &
    {{M^{(na)}+k}\choose{M^{(na)}, k}} 
  \left\{k! \prod_{m\in \calM^{(a)}}
  \e^{-u S_m}S_j^{n_m} 
   \delta_{\tau_m}(\theta^{\star}_m)
 h(S_m)p_0(\tau_m)
\right\} \nonumber \\ & & \times  
 \left\{   
  M^{(na)}! q_{M^{(na)}+k}  \prod_{m\in \calM^{(na)}}
  \e^{-u S_m}
   h(S_m)p_0(\tau_m)
    \right\} 
  \label{eq:post_3}
\end{eqnarray}
This implies that, conditionally on the auxiliary variable $U_n=u$ and on the sample $(\theta_1,\dots,\theta_n)$, 
$\tilde P$ is the superpostition of two spatial point processes.
In fact, the Jannossy density  of $\tilde P$ can be factorized as
\begin{equation}
\label{eq:condindP}
\mathcal{L}(\tilde P|u,\boldsymbol{\theta} )=
{{M^{(na)}+k}\choose{M^{(na)},k}}
\mathcal{L}(\tilde P_{a}|u,\boldsymbol{\theta} )\times 
\mathcal{L}(\tilde P^{(na)}\mid u,\boldsymbol{\theta} )
\end{equation}
We now determine the density of the two processes, the one corresponding to the
allocated jumps and the process defining the  unallocated jumps. In the first case, 
we have that:
\begin{equation*}
  \mathcal{L}(\widetilde P_a\mid \mathcal{M}^{(a)},u,\boldsymbol{\theta} ) \propto
  k!\prod_{m\in \calM^{(a)}}
  \e^{-u S_m}S_j^{n_m} \delta_{\theta^{\star}_{m}}(\tau_m) h(S_m).
\end{equation*}
 This implies that  $\widetilde P^{(a)}$ is a set of $k$ independent points.
In particular, we have 
$\widetilde P^{(a)}=\{(S_1,\theta_1^{\star}),\dots,(S_k,\theta_k^{\star})\}$, for fixed
$\theta^\star_m$s.  The $S_m$s have density given by
\begin{equation*}
  h_{n_m}(s;u)=\left( (-1)^{n_m}\frac{d}{du^{n_m}}\psi(u)\right)^{-1} 
  \e^{-u s}s^{n_m}  h(s), \quad s>0,
\end{equation*}
where 
\begin{equation*}
  \psi(u):=\int_{0}^{\infty}e^{-u s}h(s)ds,\quad u>0
\end{equation*}
is the Laplace transform of the density $h(s)$.

\noindent We now focus on the process of unallocated jumps. It is straightforward to derive the 
Janossy density of the  process $\widetilde{P}^{(na)}$:
\begin{equation}
  \label{eq:janossy_nonallocated}
  \begin{split}
    \mathcal{L}(\widetilde P^{(na)}\mid \mathcal{M}^{(na)},u,\boldsymbol{\theta} ) 
    &\propto
    (M^{(na)}+k)! q_{M^{(na)}+k}  \prod_{m\in \calM_{na}}
    \e^{-u S_m}
    h(S_m)p_0(\tau_m)\\
    &=
    \left\{ \frac{(M^{(na)}+k)!}{M^{(na)}!} \psi(u)^{M^{(na)}} q_{M^{(na)}+k} \right\} M^{(na)}!
    \prod_{m\in \calM^{(na)}}^{ } \psi(u)^{-1} \e^{-u S_m}
    h(S_m)p_0(\tau_m)\\
    &\propto \widetilde{q}_{M^{(na)}} M^{(na)}!
    \prod_{m\in \calM^{(na)}}^{ } \psi(u)^{-1} \e^{-u S_m}
    h(S_m)p_0(\tau_m)
  \end{split}
\end{equation}
where 
$$
\widetilde{q}_{M^{(na)}}=\frac{1}{C_{k}} \frac{(M^{(na)}+k)!}{M^{(na)}!} \psi(u)^{M^{(na)}}
q_{M^{(na)}+k} 
$$
Since
$$C_{k}:= \sum_{M^{(na)}=0}^{\infty}\frac{(M^{(na)}+k)!}{M^{(na)}!} \psi(u)^{M^{(na)}}q_{M^{(na)}+k}
<\sum_{M^{(na)}=0}^{\infty}\frac{(M^{(na)}+k)!}{M^{(na)}!} \psi(u)^{M^{(na)}}
=\frac{k!}{(1-\psi(u))^{k+1}}
$$
we can conclude that  $\{\widetilde{q}_{M^{(na)}}\}$ is a discrete probability on
$\{0,1,\dots\}$ and Eq.~\eqref{eq:janossy_nonallocated} defines a proper Janossy density. 
A posteriori, conditionally on $U_n$, $M^{(na)}$ is distributed as $\widetilde{q}_{M^{(na)}}$ and, given $M^{(na)}$, the jumps $S_m$s are i.i.d. from the exponentially tilted distribution defined by $h(s)e^{-uS}$.

This concludes the proof of item (a) and (b). Proof of item (c) follows directly from Eq.~\eqref{eq:condindP}.
Derivation of the posterior distribution of $U_n$ follows from the expression of the eppf obtained in Theorem \ref{thm:eppf_peps}. 
\end{proof}

\subsection{Proof of Eq.~\eqref{eq:margU}}
\label{sec:proof_marg_u}

\begin{align*}
  f_{U_n}(u,n)=\frac{u^{n-1}}{\Gamma(n)}\E(T^n \e^{-Tu})&=
  \frac{u^{n-1}}{\Gamma(n)}(-1)^n \frac{d}{du^n}\E\left( \e^{-Tu} \right)\\
  &= \frac{u^{n-1}}{\Gamma(n)}(-1)^n \frac{d}{du^n}\E\left( \E(\e^{-Tu}\mid M)
  \right)\\
  &=
\frac{u^{n-1}}{\Gamma(n)}(-1)^n \frac{d}{du^n}\E\left( \prod_{m=1}^{M} 
\E(\e^{-S_m u}) \right)\\
&=\frac{u^{n-1}}{\Gamma(n)}(-1)^n \frac{d}{du^n}\E\left( 
\psi(u)^M \right)\\
\end{align*}

\subsection{Proof of Proposition \ref{prop:bessel}}
\label{sec:app_proof_bessel}

  \begin{proof}
The Laplace transform of  $S_m$ can be rewritten via the L\'evy-Khintchine formula:
\begin{align*}
  \psi(u) &:=  \E(\e^{-\lambda T})= \exp\left\{  -\alpha \int_0^{+\infty}(1-\e^{-\lambda s}) \rho(s;\beta)ds \right\}\\
  &= \exp\left\{ - \alpha \left( \log \left( \frac{\beta+\lambda}{\beta}\right) +\sum_{m=1}^{+\infty}\frac{\Gamma(2m)}{2^{2m}(m!)^2 \beta^m} - \sum_{m=1}^{+\infty}\frac{\Gamma(2m)}{2^{2m}(m!)^2 (\beta+\lambda)^m}
\right)\right\} \\
&= \exp\left\{ - \alpha \log \left(
  \frac{\beta+\lambda+\sqrt{(\beta+\lambda)^2-1}}{\beta+\sqrt{\beta^2-1}}
\right)\right\}.\\
&=\left( \frac{\beta+\sqrt{\beta^2-1}} {\beta+u+\sqrt{(\beta+u)^2-1}} \right)^\alpha
\end{align*}

The same expression is obtained when $S_m\sim h(s)=\alpha (\beta+\sqrt{\beta^2-1})^\alpha 
\dfrac{\e^{-\beta s}}{s}I_{\alpha}(s)$, $s>0$ \citep[see][Equation (17.13.112)]{GraRyz07} and this proves both item 
(a) and (b).  Note  that, when $\beta=1$, $f_T$ is called 
\textit{Bessel function density}
\citep{feller1971}.

To prove (c), we use the definition of the function $\kappa$ and obtain that for each 
$n_j\ge 1$
\begin{align*}
  \kappa(n_j,u)&=\E(S_j^{n_j}\e^{-uS_j})=\int_{0}^{\infty}s^{n_j}\e^{-us}h(s)ds\\
  &=\alpha (\beta+\sqrt{\beta^2-1})^\alpha\int_{0}^{\infty}s^{n_j-1} \e^{-(u+\beta)s}I_{\alpha}(s)ds\\
  &=\alpha (\beta+\sqrt{\beta^2-1})^\alpha
  \sum_{m=0}^{\infty}\frac{1}{m!\Gamma(\alpha+m+1)2^{2m+\alpha}}\int_{0}^{\infty}s^{2m+\alpha+n_j-1}\e^{-(u+\beta)s}ds\\
  &=\alpha (\beta+\sqrt{\beta^2-1})^\alpha
  \sum_{m=0}^{\infty}\frac{1}{m!\Gamma(\alpha+m+1)2^{2m+\alpha}}\frac{\Gamma(2m+\alpha+n_j)}{\left(u+\beta \right)^{2m+\alpha+n_j}}\\
  & \frac{\alpha \left(  \beta+\sqrt{\beta^2-1} \right)^\alpha}{2^{\alpha}(u+\beta)^{n_j+\alpha}} \frac{\Gamma(\alpha+n_j)}{\Gamma(\alpha+1)}\ff \left(\frac{n_j+\alpha}{2}, \frac{n_j+\alpha+1}{2};\alpha+1; \frac{1}{(u+\beta)^2}\right) 
\end{align*}

\end{proof}

\subsection{Proof of Convergence of the Finite DP}
\label{app:conv}
To prove convergence of the finite Dirichlet process to the infinite DP when $\gamma =\alpha / \Lambda$ goes to infinity, we need the following results.  
\begin{itemize}

  \item Let  $\{P_t\}_{t=1,2\dots}$ be a sequence of species sampling processes defined on the same space $\Theta$ as  $P$. 
    We denote by $\pi_t(n_1,\dots,n_k)$ the eppf of $P_t$ for each $t=1,2,\dots$ and $\pi_0(n_1,\dots,n_k)$ the eppf of $P_0$.
    Then, if
    \begin{align*}
      &a)\ \pi_{t}(n_1,\dots,n_k)=\int_{0}^{\infty}\tilde\pi_t(u;n_1,\dots,n_k)du, \quad \text{for each } t\\
      &b)\ \pi_{0}(n_1,\dots,n_k)=\int_{0}^{\infty}\tilde\pi_0(u;n_1,\dots,n_k)du\\
      &c)\ \lim_{ t\rightarrow\infty} \pi_{t}(u,n_1,\dots,n_k)=\pi_{0}(u,n_1,\dots,n_k) \text{ for each } u>0\ \text{and } n_1,\dots,n_k
   \end{align*}
   then $\lim_{ t\rightarrow\infty} \pi_{t}(n_1,\dots,n_k)=\pi_{0}(n_1,\dots,n_k) \text{ for each } n_1,\dots,n_k$, moreover this latter implies that $\{P_t\}$ converges in law to $P$ \citep[see][]{argiento2016}. 
\end{itemize}
Now let $n_1,\dots,n_k$ be a composition of $n$.  the eppf of the Dirichlet process is 
\begin{equation}
 \pi_0(n_1,\dots,n_k)=\frac{\Gamma(\alpha)}{\Gamma(\alpha+n)}\prod_{j=1}^{k}\alpha\Gamma(n_j)
  \label{eq:eppf_dir1}
\end{equation}
 It is trivial to observe that
\begin{equation}
   \pi_0(n_1,\dots,n_k)=
\int_{0}^{\infty}\tilde\pi_0(u;n_1,\dots,n_k)du=
   \int_{0}^{\infty}\frac{u^{n-1}(u+1)^{-(\alpha+n)}}{\Gamma(n)}\prod_{j=1}^{k}\alpha \Gamma(n_j)du
  \label{eq:eppf_dir_wu}
\end{equation}
From Eq.~\eqref{eq:eppf_dir_wu} we see that condition $b)$ is satisfied. We now check that also condition $a)$ is satisfied.
Let $\pi_\Lambda(n_1,\dots,n_k)$ be the eppf of a Finite DP with parameter $\Lambda$ and $\gamma=\alpha/\Lambda$. From Eq.~\eqref{eq:eppf_dir1} we can easily derive 
\begin{equation*}
  \begin{split}
    \pi_\Lambda(n_1,\dots,n_k)&=\int_{0}^{\infty}\tilde{\pi}_{\Lambda}(u;n_1,\dots,n_k)du\\
    &=\int_{0}^{\infty}
    \frac{u^{n-1}}{\Gamma(n)}
    \frac{\Lambda+k(u+1)^{\alpha/\Lambda}}{\Lambda(u+1)^{\alpha(k+1)/\Lambda}(u+1)^n}
  \exp\left\{-\Lambda\frac{(u+1)^{\alpha/\Lambda}-1}{(u+1)^{\alpha/\Lambda}}\right\}\\
  &\hspace{8cm}
  \times \prod_{j=1}^{k}\Lambda\frac{\Gamma(\alpha/\Lambda+n_j)} {\Gamma(\alpha/\Lambda)}du 
\end{split}
  \label{epp:dir_Lam}
\end{equation*}
We observe now that
\begin{enumerate}[i.]
  \item 
    $$\lim_{\Lambda \rightarrow \infty}
    \frac{\Lambda+k(u+1)^{\alpha/\Lambda}}{\Lambda(u+1)^{\alpha(k+1)/\Lambda}(u+1)^n}=\frac{1}{(u+1)^n}
    $$
  \item 
    $$\lim_{\Lambda \rightarrow \infty} \exp\left\{-\Lambda\frac{(u+1)^{\alpha/\Lambda}-1}{(u+1)^{\alpha/\Lambda}}\right\}=\frac{1}{(1+u)^{\alpha}}$$
  \item For each $j=1,\dots,k$
$$\lim_{\Lambda \rightarrow \infty}\Lambda\frac{\Gamma(\alpha/\Lambda+n_j)} {\Gamma(\alpha/\Lambda)}=\alpha\Gamma(n_j)$$
%
%
\end{enumerate}
From the previous three items,  we conclude that 
\begin{equation*}
  \lim_{\Lambda \rightarrow \infty}\tilde{\pi}_{\Lambda}(u;n_1,\dots,n_k)=\tilde{\pi}_0(u;n_1,\dots,n_k)
\end{equation*}
which proves that condition $c)$ is also verified.

\section{Appendix: The conditional Gibbs sampler for Finite Dirichlet Mixture model}
\label{sec:alg_fmm}
Here we describe the conditional algorithm for Finite Mixture models in the particular case of Subsection \ref{sec:FDP}. In what follows we also place a prior on $\gamma$ and $\Lambda$ as implemented in the Example of Section \ref{sec:Galaxy}.
The full model is:
\begin{eqnarray*}
Y_i \mid c_i , \tau_1 , \ldots, \tau_M, M &\sim & f(y\mid \tau_{c_i}) \\
\tau_i\mid M &\iid & p_0(\tau)\\
\mathbf{w} \mid M &\sim & \text{Dirichlet}_M(\gamma,\cdots,\gamma)\\
c_i \mid M, \mathbf{w} &\sim & \text{Multinomial}_M(1, w_1,\ldots,  w_M) \\
\gamma &\sim & \text{Gamma}(a_1, b_1)
\\M &\sim & \text{Poisson} (\Lambda) \\
\Lambda &\sim & \text{Gamma}(a_2, b_2)
\end{eqnarray*}

We build a blocked Gibbs sampler to update blocks of parameters, which are drawn from  multivariate distributions.
In particular,  the parameters of interest are 
$(\tilde P,\mathbf{c}, U)$, where $\tilde P$ is the unnormalized finite point
process and $U$ is an auxiliary  variable introduced in  Theorem \ref{theo:Posterior}. Full conditionals can be derived for most of the parameters. 
The main steps of the algorithms are:
\begin{enumerate}
  \item[1.] \textbf{Sampling from $\boldsymbol\LL(U|{\bm Y},\mathbf{c},\widetilde P)$}: 
    by construction, conditionally on $\tilde P$, the random variable $U$
    is distributed as  Gamma with parameters $(n,T)$.
  \item[2.] \textbf{Sampling from 
      $\boldsymbol\LL(c_i \mid u, {\bm Y},\tilde P)$}: 
      each $c_i$, for $i=1,\dots,n$, has  a discrete law with support 
      $\{1,\dots,M\}$, and probabilities 
      $\mathbb{P}(c_i=m) \propto S_m f(Y_i; \tau_m)$. Le $k\leq M$ be the number of allocated components   
      after resampling the entire allocation vector $\mathbf{c}$. Rename the allocated components from 1 to $k$, so that the allocated clusters correspond to the first $k$ components in the mixture and the remaining $(M-k)$ are empty for $k<M$ .  
    \item[3.] \textbf{Sampling from $
	\boldsymbol\LL(M,\mathbf{w} |u,{\bm c}, {\bm Y})$}: 
	first we observe that conditionally on $ {\bm  c}$, the weights  
	$\mathbf{w}$ do not depend on the observations $\bm Y$.
	Therefore,  we have to sample from $\LL(M, \mathbf{w} \mid u,{\bm c})$. As stated in Theorem \ref{theo:Posterior}, we can split this step into three  sub-steps. Recall that we partition the vector $\mathbf{w}=(\mathbf{w}_a,\mathbf{w}_{na})$, where $\mathbf{w}_a=(w_1,\ldots,w_k)$ and $\mathbf{w}_{na})=(w_{k+1}, \ldots,w_M) $ correspond to the allocated and unallocated components, respectively. Moreover, in our construction $k$ is fixed and determined in Step 2 and $w_j \propto S_j$.
	\begin{enumerate}
	  \item[3.a]   \textbf{Sampling from $\boldsymbol{\mathcal{L}}( M \mid u,   \mathbf{c}, \boldsymbol{Y})$}: Note that $M=k + M^{(na)}$, where $M^{(na)}$ is the number of unallocated components. We sample $M^{(na)}$ from the discrete probability measure defined on$\{ 0, 1, 2, \ldots\}$: 
	    \begin{eqnarray*}
	      \Pr(M^{(na)}=m) &= & q^\star_m \\
	      &=& \frac{(u+1)^{\gamma}k}{(u+1)^{\gamma}k+\Lambda} \mathcal{P}_0(\Lambda/(u+1)^{\gamma}) 
	      + \frac{\Lambda}{(u+1)^{\gamma}k+\Lambda}\mathcal{P}_1(  \Lambda/(u+1)^{\gamma} )
	    \end{eqnarray*}
	    which corresponds to a two-components mixture. Here  $\mathcal{P}_i(\lambda)$ denotes the shifted Poisson pmf on $\{i,i+1,i+2,\ldots \}$ with mean $i+\lambda$, $i=0,1$.
	  \item[3.a] \textbf{Sampling from $\boldsymbol{\mathcal{L}}( \mathbf{w}^{(a)}|u,\mathbf{c}, \boldsymbol{Y})$}: the allocated process
	    is a set of independent pairs of variables $\left\{ (S_1,\tau_1),\dots (S_k,\tau_k) \right\}$ such that, for $m=1,\dots,k$:
	    \begin{eqnarray*}
	      S_m&\sim& \text{Gamma} (n_m+\gamma,u+1)\\  
	     \Prob( \tau_m=d\tau_m \mid \mathbf{c}, \mathbf{y} ) &\propto& \prod_{\{i:c_i=m\}}^{}f(y_i\mid \tau_m)p_0(\tau_m)d\tau_m
	    \end{eqnarray*}
	    where $n_m$ is the cardinality of the set $\{i:c_i=m\}$, i.e is the number of allocation variables $c_i$ such that $c_i=m$. When the density $p_0$ and the family of kernels $f(y\mid \tau)$ are conjugate, then all the full conditionals $\Prob( \tau_m=d\tau_m \mid \mathbf{c}, \mathbf{y} )$ are available in closed form, simplifying the implementation of the algorithm.

	  \item[3.c] \textbf{Sampling from  $\boldsymbol{\mathcal{L}}(\mathbf{w}^{(na)}\mid u,\mathbf{c}, \boldsymbol{Y},M_{na})$}: the weights corresponding to unallocated jumps can be sampled as follows 
	     \begin{eqnarray*}
	      S_m&\sim& \text{Gamma} (\gamma,u+1)\\  
	      \tau_m \mid \mathbf{c}, \mathbf{y} &\sim &p_0(\tau_m)
	    \end{eqnarray*}
	    for $m=k+1, \ldots,M$. 
	\end{enumerate}
\item[4.]    \textbf{Sampling from $\boldsymbol{\mathcal{L}}( \Lambda \mid u,   k , \gamma) $}: this steps involves a conjugate update:
\begin{eqnarray*}   
  \Lambda\mid \text{rest} &\sim & \frac{\psi(u)}{1+b_2}\text{Gamma}(a_2^\star+1,1-\psi(u)+b_2)+
       \frac{1-\psi(u)+b_2}{1+b_2}\text{Gamma} (a_2^\star,1-\psi(u)+b_2)\\
    a_2^\star &=& k+a_2\\     
     \psi(u)&=&\frac{1}{(u+1)^\gamma}
     \end{eqnarray*} 
    where $\psi(u)$
    is the Laplace transform of a
    Gamma$(\gamma,1)$ density.
    \item[5.]     \textbf{Sampling from $\boldsymbol{\mathcal{L}}( \gamma \mid u,   k , \mathbf{c} ,\Lambda)$}: we have to implement a Metropolis-Hasting step to sample from: 
      $$\pi(\gamma\mid \text{rest})\propto \left( \Lambda \psi(u)+k\right)\e^{\Lambda\psi(u)}\frac{1}{\psi(u)^k}\prod_{j=1}^{k}\frac{\Gamma(\gamma+n_j)}{\Gamma(\gamma)}
      $$
      as the full conditional is not available in closed form.  
    For the example of Section \ref{sec:Galaxy}, we have used a random walk as proposal distribution. Adaptive strategies can be  easily implemented to improve mixing. 
    \end{enumerate}
   We note that we could have opted for a Negative-Binomial distribution as prior on $M$, with a Beta hyper-prior on the probability of success.  The Negative-Binomial can be used to induce more sparsity. In this case step (3.b),(4) and (5) need to modified accordingly.

\section{Appendix: Additional  Figures}
\label{sec:add_figures}

\begin{figure}[h!]
  \centering
  \includegraphics[scale=0.5]{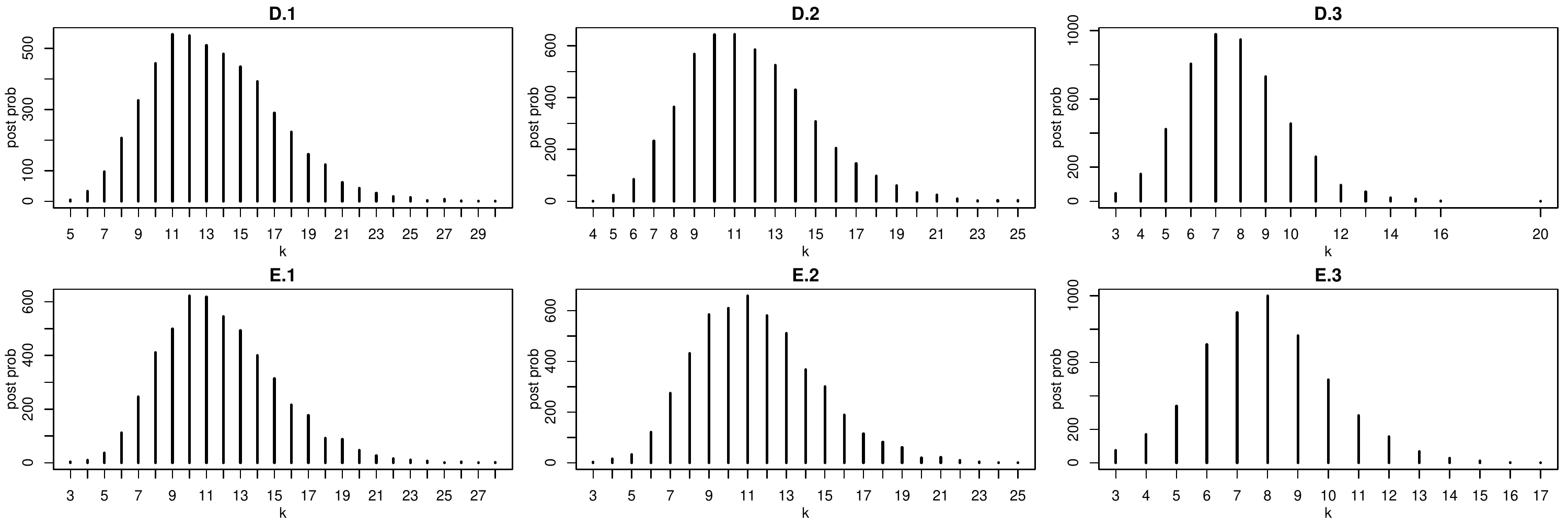}
 \includegraphics[scale=0.5]{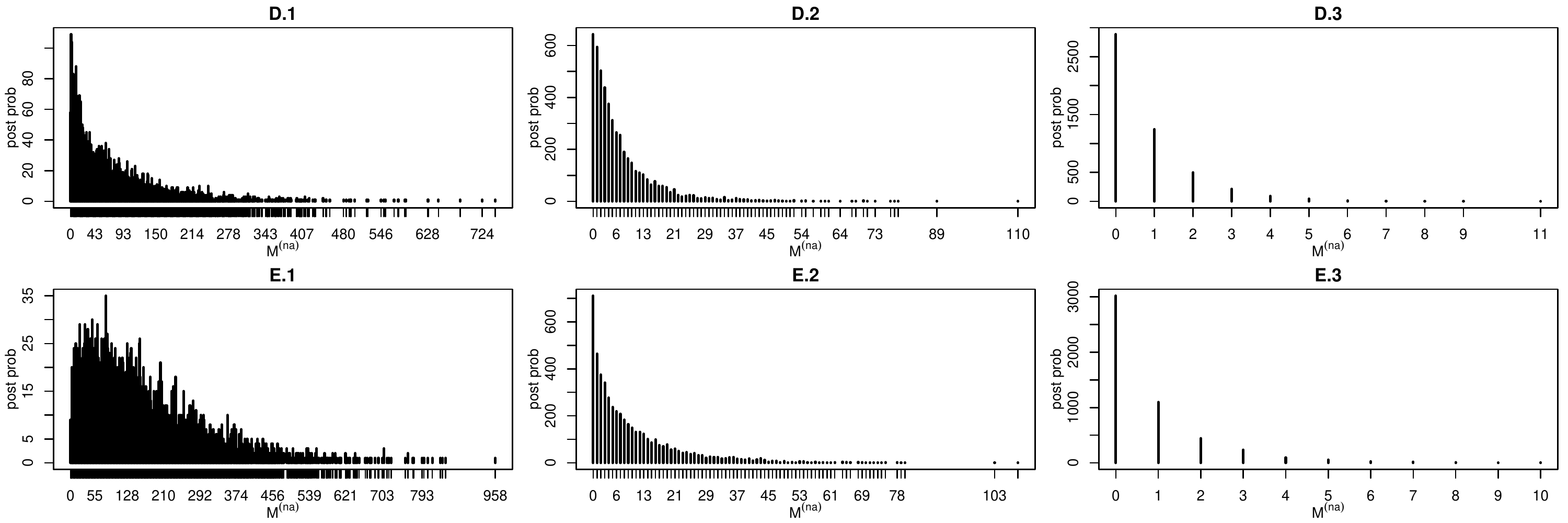}
  \caption{Posterior distribution of $k$, when $\lambda$ and $\gamma$ are random.}
  \label{fig:post_k_random}
\end{figure}
\begin{figure}[h!]
 \centering
 \includegraphics[scale=0.5]{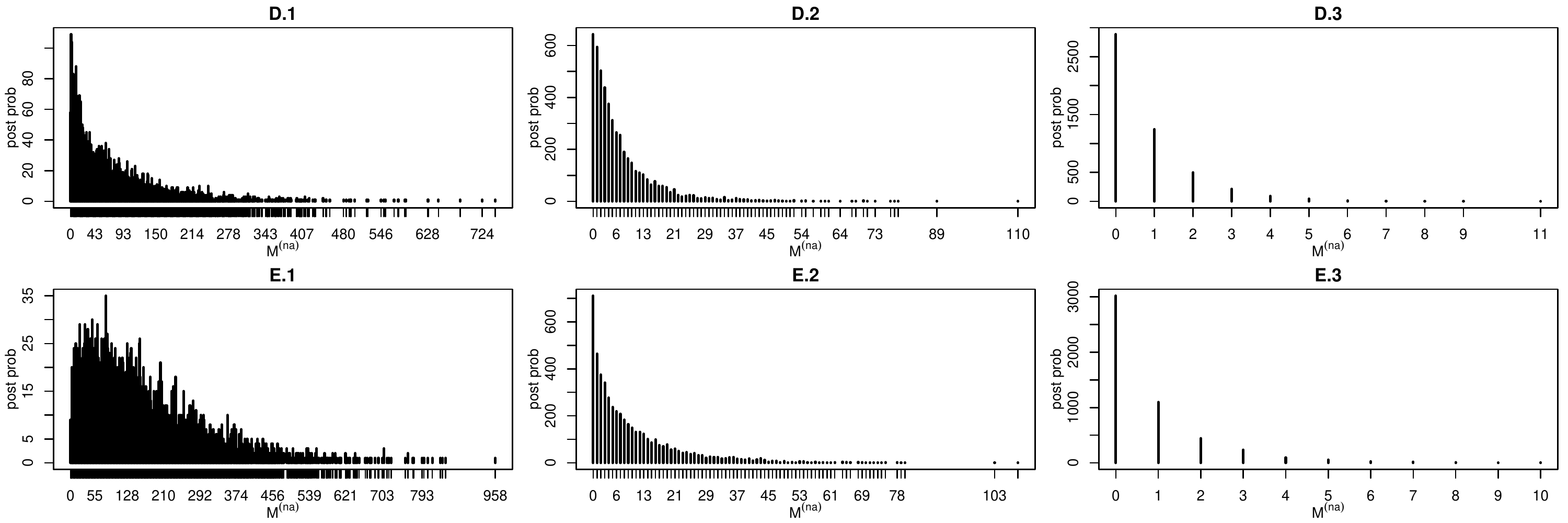}
 \caption{Posterior distribution of $M^{(na)}$, when $\lambda$ and $\gamma$ are random.}
 \label{fig:post_Mna_random}
\end{figure}

\newpage

\bibliographystyle{raf_stile}
\bibliography{bibliografia}

\begin{thebibliography}{68}
\expandafter\ifx\csname natexlab\endcsname\relax\def\natexlab#1{#1}\fi
\expandafter\ifx\csname url\endcsname\relax
  \def\url#1{\texttt{#1}}\fi
\expandafter\ifx\csname urlprefix\endcsname\relax\def\urlprefix{URL: }\fi

\bibitem[{Aldous(1985)}]{aldous1985exchangeability}
Aldous, D.~J. (1985) Exchangeability and related topics.
\newblock In \textit{{\'E}cole d'{\'E}t{\'e} de Probabilit{\'e}s de Saint-Flour
  XIII-1983}, 1--198. Springer.

\bibitem[{Argiento et~al.(2016)Argiento, Bianchini and
  Guglielmi}]{argiento2016}
Argiento, R., Bianchini, I. and Guglielmi, A. (2016) Posterior sampling from
  $\varepsilon$-approximation of normalized completely random measure mixtures.
\newblock \textit{Electron. J. Statist.}, \textbf{10}, 3516--3547.
\newblock \urlprefix\url{https://doi.org/10.1214/16-EJS1168}.

\bibitem[{Argiento et~al.(2019)Argiento, Cremaschi and
  Vannucci}]{Arg_etal18_sub}
Argiento, R., Cremaschi, A. and Vannucci, M. (2019) Hierarchical normalized
  completely random measures to cluster grouped data.
\newblock \textit{Journal of the American Statistical Association}, to appear.
\newblock
  \urlprefix\url{https://www.carloalberto.org/research/working-papers/}.

\bibitem[{Biernacki et~al.(2000)Biernacki, Celeux and
  Govaert}]{biernacki2000assessing}
Biernacki, C., Celeux, G. and Govaert, G. (2000) Assessing a mixture model for
  clustering with the integrated completed likelihood.
\newblock \textit{IEEE transactions on pattern analysis and machine
  intelligence}, \textbf{22}, 719--725.

\bibitem[{Callens et~al.(2011)Callens, Galbusera, Matthysen, Durand, Githiru,
  Huyghe and Lens}]{callens2011genetic}
Callens, T., Galbusera, P., Matthysen, E., Durand, E.~Y., Githiru, M., Huyghe,
  J.~R. and Lens, L. (2011) Genetic signature of population fragmentation
  varies with mobility in seven bird species of a fragmented kenyan cloud
  forest.
\newblock \textit{Molecular Ecology}, \textbf{20}, 1829--1844.

\bibitem[{Charalambides(2005)}]{charalambides2005combinatorial}
Charalambides, C.~A. (2005) \textit{Combinatorial methods in discrete
  distributions}, vol. 600.
\newblock John Wiley \& Sons.

\bibitem[{Daley and Vere-Jones(2007)}]{daleyverej}
Daley, D.~J. and Vere-Jones, D. (2007) \textit{An introduction to the theory of
  point processes: volume II: general theory and structure}.
\newblock Springer Science \& Business Media.

\bibitem[{Dellaportas and Papageorgiou(2006)}]{dellaportas2006multivariate}
Dellaportas, P. and Papageorgiou, I. (2006) Multivariate mixtures of normals
  with unknown number of components.
\newblock \textit{Statistics and Computing}, \textbf{16}, 57--68.

\bibitem[{DeVore and Lorentz(1993)}]{devore1993constructive}
DeVore, R.~A. and Lorentz, G.~G. (1993) \textit{Constructive approximation},
  vol. 303.
\newblock Springer Science \& Business Media.

\bibitem[{Devroye(2009)}]{devroye2009random}
Devroye, L. (2009) Random variate generation for exponentially and polynomially
  tilted stable distributions.
\newblock \textit{ACM Transactions on Modeling and Computer Simulation
  (TOMACS)}, \textbf{19}, 18.

\bibitem[{Dey et~al.(2012)Dey, M{\"u}Iler and Sinha}]{dey2012practical}
Dey, D.~D., M{\"u}Iler, P. and Sinha, D. (2012) \textit{Practical nonparametric
  and semiparametric Bayesian statistics}, vol. 133.
\newblock Springer Science \& Business Media.

\bibitem[{Eddelbuettel and Fran\c{c}ois(2011)}]{rcpp2011}
Eddelbuettel, D. and Fran\c{c}ois, R. (2011) {Rcpp}: Seamless {R} and {C++}
  integration.
\newblock \textit{Journal of Statistical Software}, \textbf{40}, 1--18.
\newblock \urlprefix\url{http://www.jstatsoft.org/v40/i08/}.

\bibitem[{Erd{\'e}lyi et~al.(1953)Erd{\'e}lyi, Magnus, Oberhettinger, Tricomi
  and Bateman}]{erdelyi53}
Erd{\'e}lyi, A., Magnus, W., Oberhettinger, F., Tricomi, F.~G. and Bateman, H.
  (1953) \textit{Higher transcendental functions}, vol.~2.
\newblock McGraw-Hill New York.

\bibitem[{Falush et~al.(2003)Falush, Stephens and
  Pritchard}]{falush2003inference}
Falush, D., Stephens, M. and Pritchard, J.~K. (2003) Inference of population
  structure using multilocus genotype data: linked loci and correlated allele
  frequencies.
\newblock \textit{Genetics}, \textbf{164}, 1567--1587.

\bibitem[{Favaro et~al.(2015)Favaro, Nipoti, Teh et~al.}]{favaro2015random}
Favaro, S., Nipoti, B., Teh, Y.~W. et~al. (2015) Random variate generation for
  {L}aguerre-type exponentially tilted $\alpha-$stable distributions.
\newblock \textit{Electronic Journal of Statistics}, \textbf{9}, 1230--1242.

\bibitem[{Favaro and Teh(2013)}]{favaro2013mcmc}
Favaro, S. and Teh, Y.~W. (2013) {MCMC} for normalized random measure mixture
  models.
\newblock \textit{Statistical Science}, 335--359.

\bibitem[{Feller(1971)}]{feller1971}
Feller, W. (1971) \textit{An Introduction to Probability Theory and Its
  Applications, vol. II}.
\newblock John Wiley, New York, second edition edn.

\bibitem[{Ferguson(1973)}]{ferguson1973bayesian}
Ferguson, T.~S. (1973) A bayesian analysis of some nonparametric problems.
\newblock \textit{The annals of statistics}, 209--230.

\bibitem[{Fr{\"u}hwirth-Schnatter(2006)}]{fruhwirth2006finite}
Fr{\"u}hwirth-Schnatter, S. (2006) \textit{Finite mixture and Markov switching
  models}.
\newblock Springer Science \& Business Media.

\bibitem[{Fruhwirth-Schnatter et~al.(2019)Fruhwirth-Schnatter, Celeux and
  Robert}]{fruhwirth2019handbook}
Fruhwirth-Schnatter, S., Celeux, G. and Robert, C.~P. (2019) \textit{Handbook
  of mixture analysis}.
\newblock Chapman and Hall/CRC.

\bibitem[{Fr{\"u}hwirth-Schnatter and Malsiner-Walli(2018)}]{fruhwirth2017here}
Fr{\"u}hwirth-Schnatter, S. and Malsiner-Walli, G. (2018) From here to
  infinity: sparse finite versus dirichlet process mixtures in model-based
  clustering.
\newblock \textit{Advances in Data Analysis and Classification}.
\newblock \urlprefix\url{https://doi.org/10.1007/s11634-018-0329-y}.

\bibitem[{Galbusera et~al.(2000)Galbusera, Lens, Schenck, Waiyaki and
  Matthysen}]{galbusera2000genetic}
Galbusera, P., Lens, L., Schenck, T., Waiyaki, E. and Matthysen, E. (2000)
  Genetic variability and gene flow in the globally, critically-endangered
  taita thrush.
\newblock \textit{Conservation Genetics}, \textbf{1}, 45--55.

\bibitem[{Geisser and Eddy(1979)}]{geisser1979predictive}
Geisser, S. and Eddy, W.~F. (1979) A predictive approach to model selection.
\newblock \textit{Journal of the American Statistical Association},
  \textbf{74}, 153--160.

\bibitem[{Gelfand and Kottas(2002)}]{gelfand2002computational}
Gelfand, A.~E. and Kottas, A. (2002) A computational approach for full
  nonparametric bayesian inference under dirichlet process mixture models.
\newblock \textit{Journal of Computational and Graphical Statistics},
  \textbf{11}, 289--305.

\bibitem[{Goutis and Robert(1998)}]{goutis1998model}
Goutis, C. and Robert, C.~P. (1998) Model choice in generalised linear models:
  A bayesian approach via kullback-leibler projections.
\newblock \textit{Biometrika}, \textbf{85}, 29--37.

\bibitem[{Gradshteyn and Ryzhik(2007)}]{GraRyz07}
Gradshteyn, I. and Ryzhik, L. (2007) \textit{Table of integrals, series, and
  products - Seventh Edition}.
\newblock San Diego (USA): Academic Press, sixth edn.

\bibitem[{Green(1995)}]{green1995reversible}
Green, P.~J. (1995) Reversible jump markov chain monte carlo computation and
  bayesian model determination.
\newblock \textit{Biometrika}, \textbf{82}, 711--732.

\bibitem[{Hofert(2011)}]{hofert2011sampling}
Hofert, M. (2011) Sampling exponentially tilted stable distributions.
\newblock \textit{ACM Transactions on Modeling and Computer Simulation
  (TOMACS)}, \textbf{22}, 3.

\bibitem[{Huelsenbeck and Andolfatto(2007)}]{huelsenbeck2007inference}
Huelsenbeck, J.~P. and Andolfatto, P. (2007) Inference of population structure
  under a dirichlet process model.
\newblock \textit{Genetics}, \textbf{175}, 1787--1802.

\bibitem[{Huelsenbeck et~al.(2011)Huelsenbeck, Andolfatto and
  Huelsenbeck}]{huelsenbeck2011structurama}
Huelsenbeck, J.~P., Andolfatto, P. and Huelsenbeck, E.~T. (2011) Structurama:
  Bayesian inference of population structure.
\newblock \textit{Evolutionary Bioinformatics}, \textbf{7}, EBO--S6761.

\bibitem[{Ishwaran and James(2001)}]{ishwaran2001gibbs}
Ishwaran, H. and James, L.~F. (2001) Gibbs sampling methods for stick-breaking
  priors.
\newblock \textit{Journal of the American Statistical Association},
  \textbf{96}, 161--173.

\bibitem[{Ishwaran and James(2003)}]{ishwaran2003some}
--- (2003) Some further developments for stick-breaking priors: finite and
  infinite clustering and classification.
\newblock \textit{Sankhy{\=a}: The Indian Journal of Statistics}, 577--592.

\bibitem[{Ishwaran and Zarepour(2002)}]{ishwaran2002exact}
Ishwaran, H. and Zarepour, M. (2002) Exact and approximate sum representations
  for the dirichlet process.
\newblock \textit{Canadian Journal of Statistics}, \textbf{30}, 269--283.

\bibitem[{Jacod and Shiryaev(2013)}]{jacod2013limit}
Jacod, J. and Shiryaev, A. (2013) \textit{Limit theorems for stochastic
  processes}, vol. 288.
\newblock Springer Science \& Business Media.

\bibitem[{James et~al.(2009)James, Lijoi and Pr{\"u}nster}]{james2009posterior}
James, L.~F., Lijoi, A. and Pr{\"u}nster, I. (2009) Posterior analysis for
  normalized random measures with independent increments.
\newblock \textit{Scandinavian Journal of Statistics}, \textbf{36}, 76--97.

\bibitem[{Jordan(2010)}]{jordan2010hierarchical}
Jordan, M.~I. (2010) Hierarchical models, nested models and completely random
  measures.
\newblock \textit{Frontiers of statistical decision making and Bayesian
  analysis: In honor of James O. Berger. New York: Springer}, 207--218.

\bibitem[{Kalli et~al.(2011)Kalli, Griffin and Walker}]{kalli2011slice}
Kalli, M., Griffin, J.~E. and Walker, S.~G. (2011) Slice sampling mixture
  models.
\newblock \textit{Statistics and computing}, \textbf{21}, 93--105.

\bibitem[{Kim et~al.(2006)Kim, Tadesse and Vannucci}]{kim2006variable}
Kim, S., Tadesse, M.~G. and Vannucci, M. (2006) Variable selection in
  clustering via dirichlet process mixture models.
\newblock \textit{Biometrika}, \textbf{93}, 877--893.

\bibitem[{Kom{\'a}rek(2009)}]{pacchettor}
Kom{\'a}rek, A. (2009) A new r package for bayesian estimation of multivariate
  normal mixtures allowing for selection of the number of components and
  interval-censored data.
\newblock \textit{Computational Statistics \& Data Analysis}, \textbf{53},
  3932--3947.

\bibitem[{Kong(1992)}]{kong1992note}
Kong, A. (1992) A note on importance sampling using standardized weights.
\newblock \textit{University of Chicago, Dept. of Statistics, Tech. Rep},
  \textbf{348}.

\bibitem[{Lau and Green(2007)}]{lau2007bayesian}
Lau, J.~W. and Green, P.~J. (2007) Bayesian model-based clustering procedures.
\newblock \textit{Journal of Computational and Graphical Statistics},
  \textbf{16}, 526--558.

\bibitem[{Lijoi et~al.(2007)Lijoi, Mena and Pr{\"u}nster}]{controlling}
Lijoi, A., Mena, R.~H. and Pr{\"u}nster, I. (2007) Controlling the
  reinforcement in bayesian non-parametric mixture models.
\newblock \textit{Journal of the Royal Statistical Society: Series B
  (Statistical Methodology)}, \textbf{69}, 715--740.

\bibitem[{Lijoi et~al.(2010)Lijoi, Pr{\"u}nster et~al.}]{lijoi2010models}
Lijoi, A., Pr{\"u}nster, I. et~al. (2010) Models beyond the dirichlet process.
\newblock \textit{Bayesian nonparametrics}, \textbf{28}, 3.

\bibitem[{Lo(1984)}]{lo84}
Lo, A.~Y. (1984) On a class of bayesian nonparametric estimates: I. density
  estimates.
\newblock \textit{The Annals of Statistics}, 351--357.

\bibitem[{Malsiner-Walli et~al.(2016)Malsiner-Walli, Fr{\"u}hwirth-Schnatter
  and Gr{\"u}n}]{malsiner2016model}
Malsiner-Walli, G., Fr{\"u}hwirth-Schnatter, S. and Gr{\"u}n, B. (2016)
  Model-based clustering based on sparse finite gaussian mixtures.
\newblock \textit{Statistics and computing}, \textbf{26}, 303--324.

\bibitem[{Malsiner-Walli et~al.(2017)Malsiner-Walli, Fr{\"u}hwirth-Schnatter
  and Gr{\"u}n}]{malsiner2017identifying}
--- (2017) Identifying mixtures of mixtures using bayesian estimation.
\newblock \textit{Journal of Computational and Graphical Statistics},
  \textbf{26}, 285--295.

\bibitem[{McLachlan et~al.(2000)McLachlan, Lee and
  Rathnayake}]{mclachlan2000finite}
McLachlan, G.~J., Lee, S.~X. and Rathnayake, S.~I. (2000) Finite mixture
  models.
\newblock \textit{Annual Review of Statistics and Its Application}.

\bibitem[{Miller and Harrison(2013)}]{miller2013simple}
Miller, J.~W. and Harrison, M.~T. (2013) A simple example of dirichlet process
  mixture inconsistency for the number of components.
\newblock In \textit{Advances in neural information processing systems},
  199--206.

\bibitem[{Miller and Harrison(2018)}]{miller2017mixture}
--- (2018) Mixture models with a prior on the number of components.
\newblock \textit{Journal of the American Statistical Association},
  \textbf{113}, 340--356.

\bibitem[{M{\o}ller and Waagepetersen(2003)}]{moller2003statistical}
M{\o}ller, J. and Waagepetersen, R.~P. (2003) \textit{Statistical inference and
  simulation for spatial point processes}.
\newblock Chapman and Hall/CRC.

\bibitem[{Muliere and Tardella(1998)}]{muliere1998approximating}
Muliere, P. and Tardella, L. (1998) Approximating distributions of random
  functionals of ferguson-dirichlet priors.
\newblock \textit{Canadian Journal of Statistics}, \textbf{26}, 283--297.

\bibitem[{Neal(2000)}]{neal2000markov}
Neal, R.~M. (2000) Markov chain sampling methods for dirichlet process mixture
  models.
\newblock \textit{Journal of Computational and Graphical Statistics},
  \textbf{9}, 249--265.

\bibitem[{Nobile(1994)}]{nobile94}
Nobile, A. (1994) \textit{Bayesian Analysis of Finite Mixture Distributions}.
\newblock Ph.D. thesis, Department of Statistics, Carnegie Mellon University.

\bibitem[{Nobile et~al.(2004)}]{nobile2004posterior}
Nobile, A. et~al. (2004) On the posterior distribution of the number of
  components in a finite mixture.
\newblock \textit{The Annals of Statistics}, \textbf{32}, 2044--2073.

\bibitem[{Pitman(1996)}]{pitman96}
Pitman, J. (1996) Blackwell-macqueen urn scheme.
\newblock \textit{Statistics, Probability, and Game Theory: Papers in Honor of
  David Blackwell}, \textbf{30}, 245.

\bibitem[{Pitman(2003)}]{pitman2003poisson}
--- (2003) Poisson-{Kingman} {P}artitions.
\newblock In \textit{Science and Statistics: a Festschrift for Terry Speed},
  vol.~40 of \textit{IMS Lecture Notes-Monograph Series}, 1--34. Hayward (USA):
  Institute of Mathematical Statistics.

\bibitem[{Pitman(2006)}]{pitman2006combinatorial}
--- (2006) Combinatorial stochastic processes.
\newblock In \textit{{\'E}cole d'{\'E}t{\'e} de Probabilit{\'e}s de Saint-Flour
  XXXII-2002}, 1--255. Springer.

\bibitem[{Pollard(1946)}]{pollard46}
Pollard, H. (1946) The representation of $\e^{-x^{\lambda}}$ as {L}aplace
  integral.
\newblock \textit{Bulletin of the American Mathematical Society}, \textbf{52},
  908--910.

\bibitem[{Pritchard et~al.(2000)Pritchard, Stephens and
  Donnelly}]{pritchard2000inference}
Pritchard, J.~K., Stephens, M. and Donnelly, P. (2000) Inference of population
  structure using multilocus genotype data.
\newblock \textit{Genetics}, \textbf{155}, 945--959.

\bibitem[{Pritchard and Wen(2003)}]{pritchard2003documentation}
Pritchard, J.~K. and Wen, W. (2003) \textit{Documentation for STRUCTURE
  software: Version 2.3.X}.
\newblock Available at
  \url{https://web.stanford.edu/group/pritchardlab/structure.html}.

\bibitem[{Regazzini et~al.(2003)Regazzini, Lijoi and
  Pr{\"u}nster}]{regazzini2003distributional}
Regazzini, E., Lijoi, A. and Pr{\"u}nster, I. (2003) Distributional results for
  means of normalized random measures with independent increments.
\newblock \textit{Annals of Statistics}, 560--585.

\bibitem[{Richardson and Green(1997)}]{richgreen97}
Richardson, S. and Green, P.~J. (1997) On bayesian analysis of mixtures with an
  unknown number of components.
\newblock \textit{Journal of the Royal Statistical Society: Series B
  (Statistical Methodology)}, \textbf{59}, 731--792.

\bibitem[{Roeder(1990)}]{roeder1990density}
Roeder, K. (1990) Density estimation with confidence sets exemplified by
  superclusters and voids in the galaxies.
\newblock \textit{Journal of the American Statistical Association},
  \textbf{85}, 617--624.

\bibitem[{Rousseau and Mengersen(2011)}]{rousseau2011asymptotic}
Rousseau, J. and Mengersen, K. (2011) Asymptotic behaviour of the posterior
  distribution in overfitted mixture models.
\newblock \textit{Journal of the Royal Statistical Society: Series B
  (Statistical Methodology)}, \textbf{73}, 689--710.

\bibitem[{Sokal(1997)}]{sokal1997monte}
Sokal, A. (1997) Monte carlo methods in statistical mechanics: foundations and
  new algorithms.
\newblock In \textit{Functional integration}, 131--192. Springer.

\bibitem[{Stephens(2000)}]{steph00}
Stephens, M. (2000) Bayesian analysis of mixture models with an unknown number
  of components-an alternative to reversible jump methods.
\newblock \textit{Annals of statistics}, 40--74.

\bibitem[{Tadesse et~al.(2005)Tadesse, Sha and Vannucci}]{tadesse2005bayesian}
Tadesse, M.~G., Sha, N. and Vannucci, M. (2005) Bayesian variable selection in
  clustering high-dimensional data.
\newblock \textit{Journal of the American Statistical Association},
  \textbf{100}, 602--617.

\bibitem[{Wiper et~al.(2001)Wiper, Insua and Ruggeri}]{wiper2001mixtures}
Wiper, M., Insua, D.~R. and Ruggeri, F. (2001) Mixtures of gamma distributions
  with applications.
\newblock \textit{Journal of Computational and Graphical Statistics},
  \textbf{10}, 440--454.

\end{thebibliography}

\end{document}